\documentclass[a4paper,12pt]{article}
\usepackage{epsfig}
\usepackage{amsmath}
\usepackage{amssymb}

\def\today{\number\day
	   \space\ifcase\month\or
	     January\or February\or March\or April\or May\or June\or
	     July\or August\or September\or October\or November\or December\fi
	   \space\number\year}
\def\thisday{July 26, 2000}
\def\beq{\begin{equation}}
\def\eeq{\end{equation}}
\def\bea{\begin{eqnarray}}
\def\eea{\end{eqnarray}}

\def\ra{\rightarrow}

\def\tw{$t \, W^-~$}

\begin{document}

\thispagestyle{empty}
\hspace*{-0.8cm} {\vspace*{-0.5cm} \thisday}  
\vbox{ \begin{flushright}
hep-ph/0007298~~~~~~~~~~ \\ ANL-HEP-PR-98-124~~~~~~~~~~ \\ 
CERN-TH/2000-224 ~~~~~~~~~~\end{flushright} }

\vspace{1.5cm} 

\begin{center}
{ {\Large {\bf Single Top Production as a Window to Physics \\[0.4cm]
Beyond the Standard Model}}}

\vspace{1.5cm} 

\centerline{ Tim M. P. Tait }
\vspace{0.2cm}
{\it Argonne National Laboratory \\
Argonne, Illinois 60439, USA \\[0.2cm]
{\tt tait@anl.gov} } \\
\vspace{0.3cm}
\centerline{ and }
\vspace{0.3cm}
\centerline{ C.--P. Yuan\footnote{On leave from
Department of Physics and Astronomy, Michigan State University 
East Lansing, Michigan 48824, USA.} }
\vspace{0.2cm}
{\it Theory Division, CERN \\
CH-1211, Geneva, Switzerland \\[0.2cm]
{\tt yuan@pa.msu.edu} } \\

\vspace{1.0cm}

\begin{abstract}
\noindent 
Production of single top quarks at a high energy hadron collider is studied
as a means to identify physics beyond the standard model related to the
electroweak symmetry breaking.  The sensitivity of the $s$-channel 
$W^*$ mode, the $t$-channel $W$-gluon fusion mode, and the \tw mode to
various possible forms of new physics is assessed, and it is found that
the three modes are sensitive to different forms of new physics,
indicating that they provide complimentary information about the properties
of the top quark.  Polarization observables are also considered, and found
to provide potentially useful information about the structure of the
interactions of top.
\\[0.4cm]
\end{abstract}
\end{center}

\newpage
\setcounter{footnote}{0}
\renewcommand{\thefootnote}{\arabic{footnote}}

\section{Introduction}
\indent \indent
\label{intro}

With the discovery of the top quark at the Fermilab Tevatron 
\cite{topdisc}, the third generation of fermions in the Standard
Model (SM) is complete.  However, the question remains, is the top
quark ``just another quark?'' or ``is it something more?''.  Top is the
only quark to have a mass on the same order as the electroweak symmetry
breaking (EWSB) scale, $v \sim 246$ GeV, whereas all other 
observed fermions have masses that are a tiny fraction of this energy.
This enormous mass may be a clue that the top quark plays a special role
in the EWSB. Following that line of thinking, many of the proposed 
extensions of the SM explain the large top mass by allowing the top 
to participate in new dynamics 
\cite{topcondensate,topcolor,topflavor,tf1,tf2}, 
which is connected to the physics 
providing the mass of the $W$ and $Z$ bosons.
Thus, one of the primary motivations for the Tevatron Run~II and CERN
LHC\footnote{The Tevatron Run~II is a proton-anti-proton
collider with $\sqrt{S} = 2$ TeV, and the LHC is a proton-proton
collider with $\sqrt{S} = 14$ TeV.}
is to accurately determine the top quark's properties, to see
if any hint of nonstandard physics may be visible, and thus provide
information about the mechanism of mass generation 
\cite{mythesis,LHCreport}.

Top quarks may be produced in pairs at a hadron collider via the
strong interaction, through processes such as 
$q \bar{q} \ra t \bar{t}$
and $g g \ra t \bar{t}$.  Thus, the rate and kinematic distributions of
top quarks produced in this way are a measure of the top's interactions with
the gluons.  The top decay proceeds via the weak interaction, and as we
shall see does provide interesting information about the chiral structure
of the $W$-$t$-$b$ interaction \cite{mythesis,adlernelson}.  
However, decays are experimentally
relatively insensitive to the magnitude of the interaction by which they 
are mediated.  For example, in the case of top, there is one SM decay mode,
$t \ra b \, W^+$.  If this vertex were somehow modified by new physics to
have a different magnitude, it would affect the top's intrinsic width.
However, at a hadron collider the width cannot be measured because the
experimental resolutions are much larger than the width itself
\cite{mrenna}.
Similarly, while observing exotic top decays would certainly be interesting
and would suggest what type of new vertices describe the observed decays,
it would not determine the magnitude of these new interactions.  Even a
study of branching fractions compared to the SM decay mode 
may be misleading, because one must  have already measured the 
$W$-$t$-$b$ interaction strength itself through some other means.

These drawbacks lead one to study weak production mechanisms
of the top quark,
which have cross sections directly proportional to the
top's weak couplings.  The $Z$-$t$-$t$ coupling will presumably
contribute to $t \bar{t}$ production, though distinguishing this
from the much larger QCD $t \bar{t}$ production is most likely
hopeless at hadron colliders. (Nevertheless, it can be precisely measured 
at an electron Linear Collider (LC) \cite{nlcztt}.)
A better process to measure this interaction is the production of
$t \, \bar{t} \, Z$, where the $Z$ can be observed directly.
The $W$-$t$-$b$ interaction will allow one to produce single top quarks
at a hadron collider, and thus directly measure the properties of this
interaction.
A further consideration is that new physics
characterized by energy scale $\Lambda$ may be more apparent
in higher energy processes, closer to $\Lambda$.  Thus, 
new physics contributions to single top
production would scale as $(\sqrt{s} / \Lambda)^n$ (where $s$ is the
average center of mass energy of the reaction and $n$ is a
positive integer or zero) whereas top
decay processes scale as $(m_t / \Lambda)^n$.  At high energy
colliders such as the Tevatron and LHC, $\sqrt{s}$ can be
considerably larger than $m_t$, thus enhancing the relative
importance of new physics in single top production.

\begin{figure}[t]
\epsfysize=1.8in
\centerline{\epsfbox{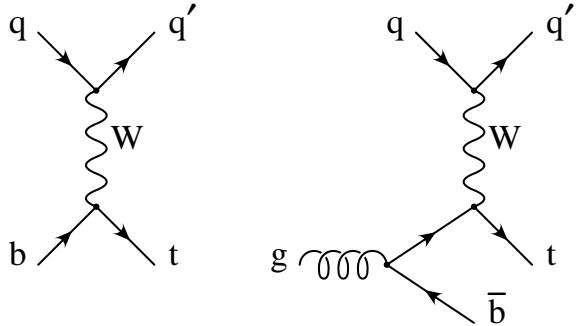}}
\caption{Feynman diagrams for $t$-channel production of
a single top in the SM.}
\label{tchanfig}
\end{figure}

\begin{figure}[t]
\epsfysize=1.25in
\centerline{\epsfbox{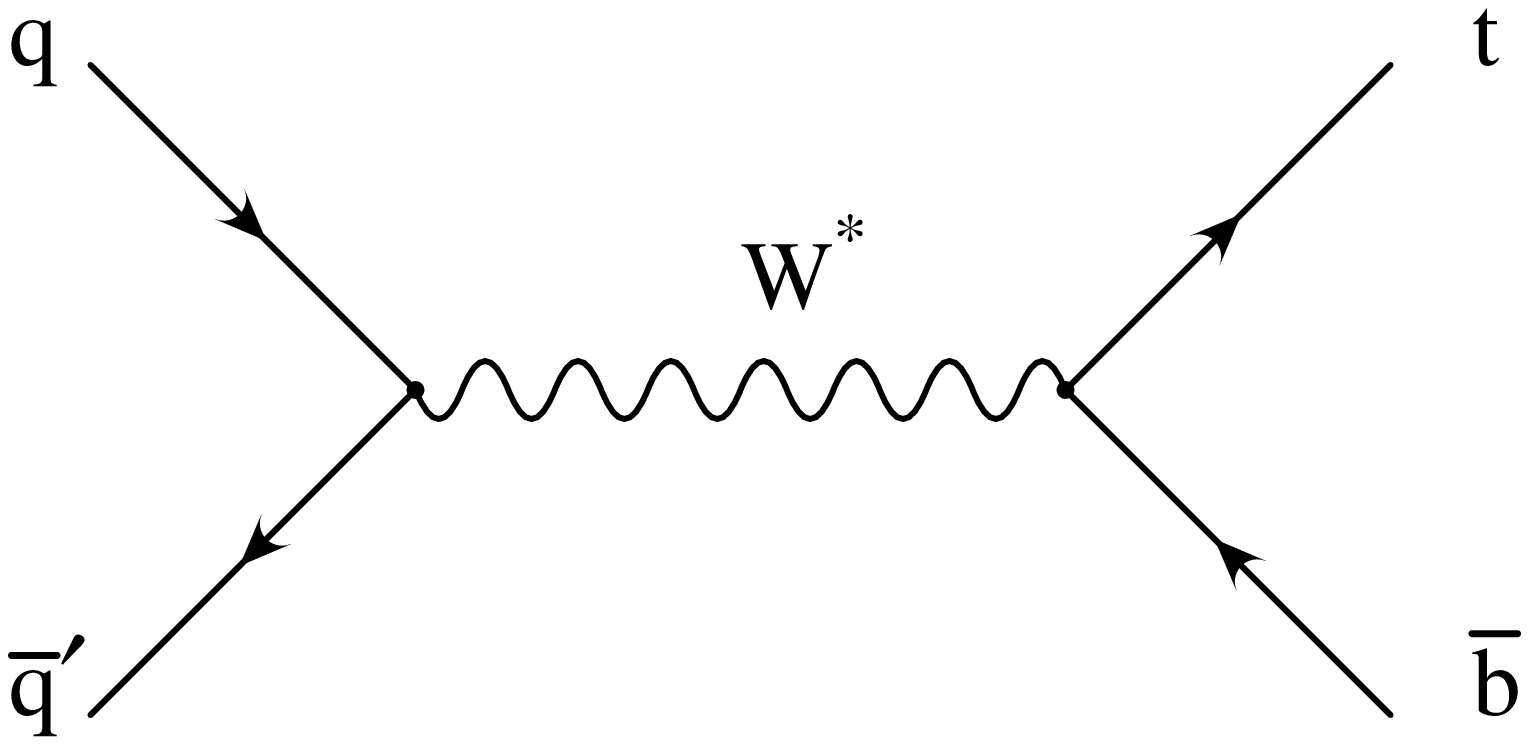}}
\caption{Feynman diagram for $s$-channel production of
a single top in the SM.}
\label{schanfig}
\end{figure}

\begin{figure}[t]
\epsfysize=1.5in
\centerline{\epsfbox{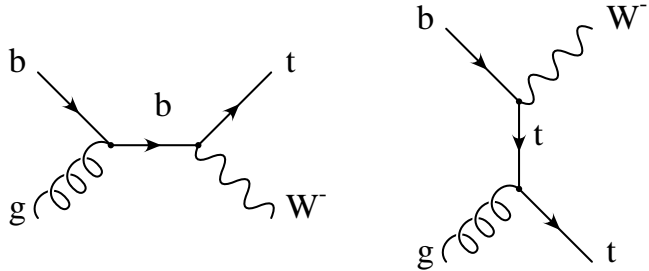}}
\caption{Feynman diagrams for \tw production in the SM.}
\label{twfig}
\end{figure}

Single top production proceeds through three distinct
sub-processes at a hadron collider.  
The $t$-channel $W$-gluon fusion
mode \cite{tchannel,dthesis,newt}
involves the exchange of a space-like $W$ boson between a
light quark, and a bottom ($b$) quark inside the incident hadrons,
resulting in a jet and a single top quark (c.f. Figure~\ref{tchanfig}).
Its rate\footnote{In quoting
cross sections, we use a top mass of 175 GeV and CTEQ4M (for $s$- and $t$-
channel processes) and CTEQ4L (for the \tw mode) parton densities
\cite{cteq4} with scales chosen as 
explained in \cite{mythesis}, where
other choices of parton densities, scale, and top mass are also
examined.  The rates of $t$ and $\bar{t}$ are summed unless otherwise
noted.}
is rather large at both the Tevatron (2.4 pb) and the LHC (243 pb).
The $s$-channel
$W^*$ mode \cite{schannel}
involves production of an off-shell, time-like $W$ boson,
which then decays into a top and a bottom quark
(c.f. Figure~\ref{schanfig}).  
It has a relatively
large rate at the Tevatron (0.86 pb), 
but is comparatively small at the LHC (11 pb) because
it is driven by initial state anti-quark parton densities.
Finally, the \tw mode \cite{tw,mytw}
of single top production involves an initial
state $b$ quark emitting a (close to) on-shell $W^-$ boson,
resulting in a \tw final state (c.f. Figure~\ref{twfig}).  
Because of the massive particles
in the final state, this mode has an extremely small rate at the 
Tevatron (0.088 pb),
but is considerable at the LHC (51 pb)
where more partonic energy is available.
Each mode has rather distinct event kinematics, and thus are
potentially observable separately from each other 
\cite{dthesis,newt,mytw}.

As we consider various new physics effects in single top production below,
it will be important to keep in mind how accurately these rates will be
measured at the Tevatron Run~II and LHC.  At the Tevatron, the combined
statistical and theoretical uncertainties for the $s$- and $t$- channel
processes will be on the order of $10-20\%$ \cite{new}, and 
the \tw rate will most likely be too small to measure\footnote{These
estimates do not include systematic uncertainties, which are beyond the
scope of this work.}.  At the LHC, the statistical uncertainties will
be close to zero, and the theoretical uncertainties on the order of
$5-10\%$ for the $s$- and $t$-channel processes \cite{new}, 
and on the order of $20\%$ for the \tw process \cite{mytw}.
However, systematic uncertainties (including the efficiency for
separating the signal from the backgrounds) are likely to
dominate these LHC estimates.

As we will see below, the three modes of single top production are
sensitive to quite different manifestations of physics beyond the
standard model.  Thus, taken together, they are a comprehensive
probe of the top quark's interactions.
In this article, we analyze several
possible signals of new physics that could manifest themselves in single top
production.  These signals can be classified as to whether they involve
the effects of a new particle (either fundamental or composite) that couple
to the top quark, or the effect of a modification of the SM coupling between
the top and other known particles.  These two classifications
can be seen to overlap in the limit in which the additional particles
are heavy and decouple from the low energy description.  In this case
the extra particles are best seen through their effects on the couplings
of the known particles.  One particular aspect of both of these sets
of new physics is the possibility of CP violation in top observables.
We will not address this issue in detail; the interested reader is
referred to the recent review \cite{cpviol} on this subject.

This article is organized as follows; in 
Section~\ref{extrap}, we investigate what sort of particles beyond those
in the SM could contribute to single top production
at a hadron collider, and what effects could be seen.  In 
Section~\ref{modint}, we study the effects of non-standard top
interactions.  In Section~\ref{polarization}, we examine how one may
use polarization observables to further learn about nonstandard physics
in the top sector.  In Section~\ref{conclusions} we summarize the picture
which emerges as the way in which the properties of top may be 
systematically extracted from the available observables.

\section{Additional Nonstandard Particles}
\label{extrap}
\indent \indent

Many theories of physics beyond the SM predict the existence of particles
beyond those required by the SM itself.  Examples include both the fundamental
superpartners in a theory with supersymmetry
(SUSY) \cite{susy}, and the composite Higgs bosons found in
top-condensation and top-color models 
\cite{topcondensate,topcolor,topflavor,tf1,tf2}.  
In order for some kind of additional
particle to contribute to single top production at tree level\footnote{
New particles may also contribute to single top production through loops
\cite{genstnp}, though the effects are generally small enough
that they are difficult to observe at a hadron collider.}
at a hadron collider, the new
particle must somehow couple the top to one of the lighter SM particles.
Thus, the new particle may be either a boson (such as a
$W^\prime$ vector boson that couples to top and bottom) 
or a fermion (such as a $b^\prime$ quark that couples to the 
$W$ boson and top).

Additional fermionic particles can couple the top and either one of
the gauge bosons or the Higgs boson.  In order to respect the 
color symmetry, this requires that the extra fermion occurs in a 
color triplet, and thus it is sensible to think of it as some
type of quark.  In order to be invariant under the electromagnetic symmetry,
this new ``quark'' should have either electric charge ($Q$)
$+2/3$ or $-1/3$ so that
one may construct gauge invariant
interactions between the extra quark, the
top quark, and the known bosons.  Generally, we can refer to a
$Q = +2/3$ extra quark as a $t^\prime$ and a $Q = -1/3$ extra quark
as a $b^\prime$, though this does not necessarily imply that the
extra quarks are in the same representation under 
${\rm SU(2)}_L \times {\rm U(1)}_Y$ as the SM top and bottom.
For example, in models where the top mass is generated by invoking
a seesaw mechanism, there are generally either SU(2) 
singlet \cite{topcolor} or doublet \cite{tf2}
quarks present in the theory.
Additional fermions are not generally expected to be
a large source of new contributions to single top production,
because of strong constraints from other 
low energy observables \cite{pdg}.  On the other
hand we will see that there are models with additional fermions
to which single top production is a sensitive probe.

Extra bosons can contribute to single top production
either by coupling top to the down-type quarks, in which case the boson
must have electric charge $Q = \pm 1$ in order to maintain the 
electromagnetic symmetry, or by coupling top to the charm or up quarks, in
which case the boson should be electrically neutral.  
There is also the possibility of a boson carrying a combination of
color and electric charge that allows it to couple top to one of the
lepton fields (this boson would then carry both lepton and baryon
quantum numbers and thus may be labelled a ``leptoquark'').
One theoretically well motivated example of a leptoquark are the gauge
bosons corresponding to the generators of a grand unified theory (GUT)
linking the SU(2) and SU(3) sectors of the SM.
This GUT picture has the leptoquark as part of the gauge interactions,
so the question as to whether or not top observables are
an interesting means to study leptoquarks becomes a question
as to whether or not the leptoquark has some reason to prefer to couple to
the top quark.  These bosons would be expected to have GUT scale masses
which in traditional GUT theories is far too large to be interesting from the
point of view of colliders envisioned in the near future\footnote{
However, a GUT theory in 5 space-time dimensions could unify at the TeV
scale \cite{tevgut}
and would be accessible to the current generation of hadron
colliders through a number of observables.}.

Another interesting picture of leptoquarks
is one in which the SM quarks and leptons are bound states of some more
fundamental set of particles (preons).
In that case the question as to whether or not the top quark is a good place
to look for evidence of the preons depends on how the model arranges the
various types of preons to build quarks and leptons.
However, at a hadron collider the possible
light parton initial states available are not suitable for production of
a single leptoquark, and thus are not particularly 
interesting in the context of single top production\footnote{
It is interesting to note that a leptoquark ($L$) with
$Q = +2 / 3$ could play an important role in top decays through a process
such as $t \ra \nu \, L \ra \nu \, b \, \ell^+$
\cite{leptoquark}.  This leads to a final
state that is identical to a SM top decay, but with a
very distinct kinematic structure.}.  For this reason,
we will not focus on leptoquarks in the discussion below.

\subsection{Extra Quarks}
\indent \indent

A simple extension of the SM is to allow for an extra set of quarks.
Such objects exist in a wide variety of extensions to the SM.
Examples of such theories include
the top seesaw versions of the top-color \cite{topcolor}
and top-flavor \cite{tf2} models, which rely on additional
fermions to participate in a seesaw mechanism to generate the top mass;
SUSY theories with gauge mediated SUSY breaking \cite{gmsb}
that must be communicated
from a hidden sector in which SUSY is broken to the visible sector
through the interactions of a set of fields with SM gauge 
quantum numbers; 
and even models with a fourth generation of fermions.

Direct search limits for QCD production of extra
quarks (at the Tevatron, for example)
require that they be quite massive ($m_{q^\prime} \ge 46 - 128$
GeV at the $95\%$ C. L., depending on the decay mode \cite{pdg}),
and thus they cannot significantly affect the single top 
production rate.
They are best observed either through their mixings
with the third family (and thus their effect on
the top couplings), or through direct production.

As a particular example, a fourth generation of quarks could
mix with the third generation through a generalized
Cabibbo-Kobayashi-Maskawa (CKM) matrix,
and this could allow $V_{tb}$ to deviate considerably from unity.
In this case, all three modes of single top production would be
expected to have considerably lower cross sections than the SM 
predicts.  This already shows how the separate modes of single top
production can be used to learn about physics beyond the SM.  Other
types of new physics could scale the three rates
independently.  Thus, if all three modes are measured to have cross
sections that are the same fraction of the SM rates, it is an
indication that the new physics modifies the top's coupling to the
bottom and $W$ (and not another pair of light particles), and
further that the modification is the same regardless of the momentum
flowing through the vertex (as is the case with the $W$-$t$-$b$
interaction in the SM).

\begin{figure}[t]
\epsfysize=1.5in
\centerline{\epsfbox{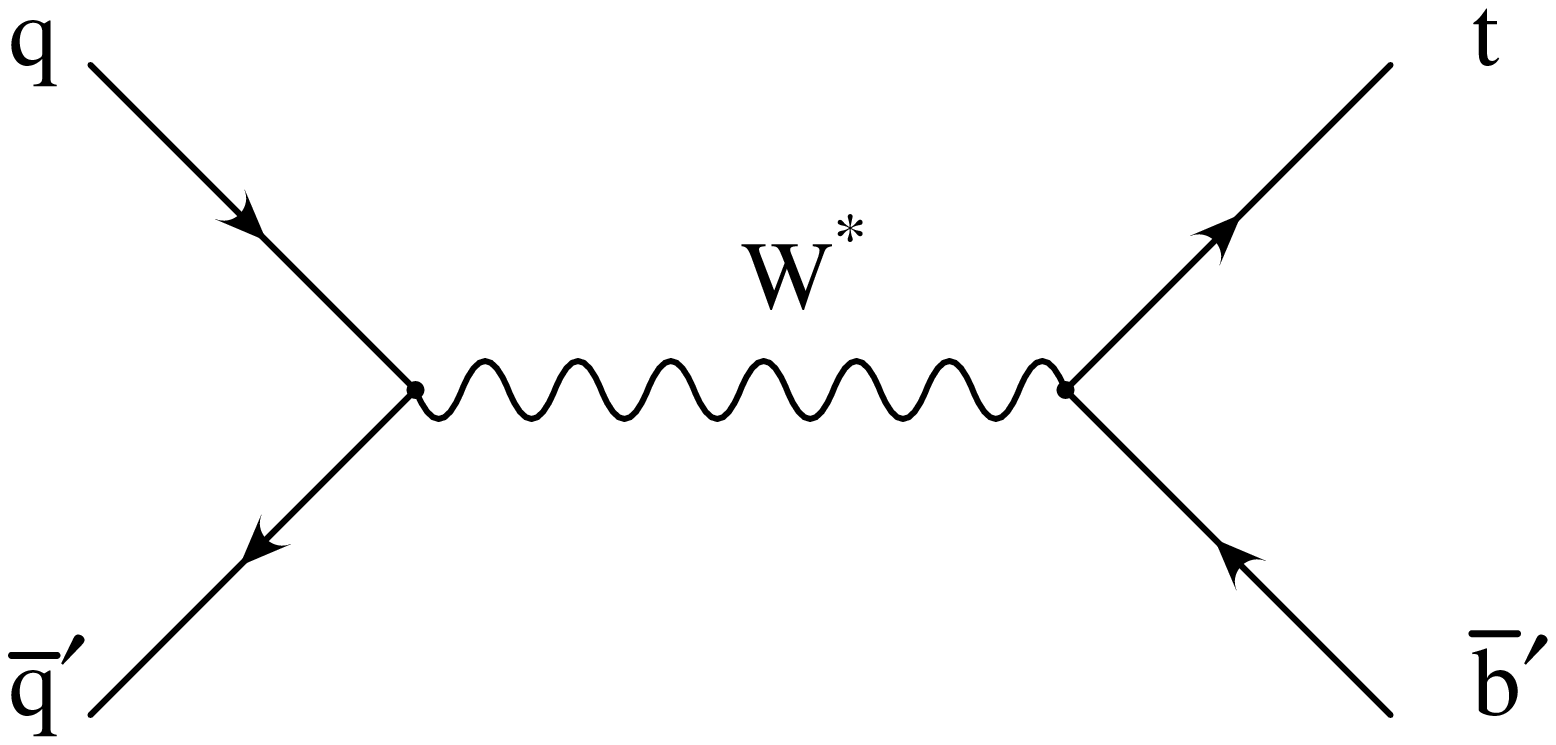}}
\caption{Feynman diagram for $s$-channel production of
a single top and a $b^\prime$:
$q \, \bar{q}^\prime \ra t \, \bar{b}^\prime$.}
\label{tbprimefig}
\end{figure}

In fact, it is instructive to examine the experimental constraints on the
CKM matrix when one does not impose three family unitarity.  The 90\%
C.L. direct constraints become \cite{pdg},
\bea
V &=&
\left(
\begin{array}{cccc}
0.9722-0.9748 & 0.216-0.223   & 0.002-0.005 & ... \\
0.199-0.233   & 0.784-0.976   & 0.037-0.043 & ... \\
0-0.09        & 0.0-0.55      & 0.06-0.9993 & ... \\
...           & ...           & ...         & ...
\end{array}
\right) ,
\eea
with two striking differences from the 3 generation matrix;
$V_{tb}$ may be significantly smaller than unity (as
mentioned above) but also that $V_{ts}$ may be as large as 0.55.
(For comparison, the corresponding limit on
$V_{ts}$ with 3 generations is $0.035 \leq |V_{ts}| \leq 0.043$
\cite{pdg}).  $|V_{ts}|= 0.043$ has a negligible effect on the 
$t$-channel rate of single top production, contributing much less
than $1\%$ of the total rate at either Tevatron or LHC.
However, if $|V_{ts}| = 0.55$ and $|V_{tb}| = 0.835$ (which saturates
the unitarity requirement for any number of families\footnote{This
choice of $V_{tb}$ and $V_{ts}$ is consistent with the CDF measurement
$\Gamma (t \ra W^+ \, b) / [\Gamma (t \ra W^+ \, b) 
+ \Gamma (t \ra W^+ \, s) + \Gamma (t \ra W^+ \, d) ] = 
0.87^{+0.13 \, +0.13}_{-0.30 \, -0.11}$
\cite{CDFvtb}.}), the 
NLO $t$-channel rate will rise to about 4.07 pb at the Tevatron
Run II and 334 pb at the LHC, thanks to the much larger $s$ quark
parton density compared to the $b$ quark.  These are huge deviations from
the SM rates, and would be a clear indication of new physics.
Under these conditions, the $s$-channel rate, which would fall to
$|V_{tb}|^2  \sim 0.7$ of its SM rate. 
Similarly to the $t$-channel mode, the \tw rate
will be enhanced to 0.19 pb at the Tevatron
and 78.1 pb at the LHC under these conditions.

\begin{figure}[p]
\epsfysize=5.0in
\centerline{\epsfbox{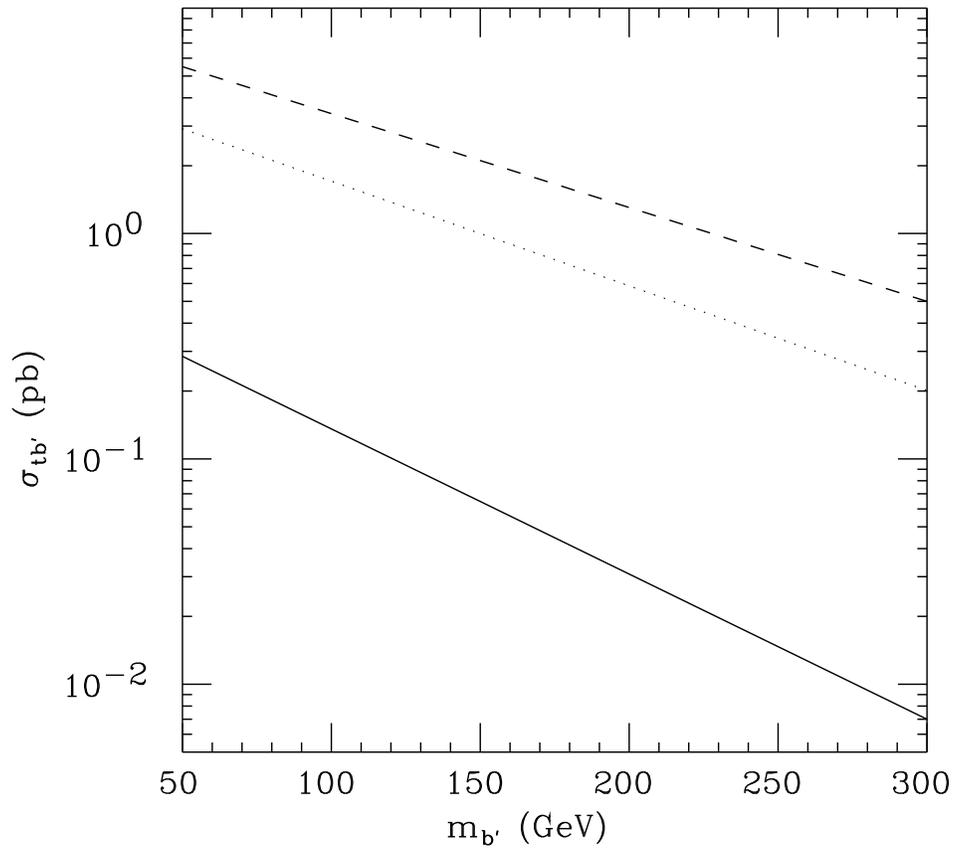}}
\caption{The NLO rates (in pb) for the process 
$q \, \bar{q}^\prime \ra W^* \ra t \, \bar{b}^\prime$ 
for various $b^\prime$
masses at the Tevatron (solid curve) and LHC (dashed curve),
assuming $V_{tb^\prime}= 1$.  At the Tevatron, the rates of
$q \, \bar{q}^\prime \ra W^* \ra \bar{t} \, b^\prime$ is equal to the
$t \, \bar{b}^\prime$ rate.  The $\bar{t} \, b^\prime$ rate at the
LHC is shown as the dotted curve.}
\label{bprimeratefig}
\end{figure}

In addition to mixing effects, one could also hope to observe direct
production of one of the fourth generation quarks, through reactions
such as $q \, \bar{q}^\prime \ra t \, \bar{b^\prime}$, shown in
Figure~\ref{tbprimefig}.  This process would provide further information
on the family structure of the 4 generation model ($V_{t b^\prime}$)
not readily 
available from the QCD production of $b^\prime \bar{b}^\prime$.
The production rates will depend on the
magnitude of the $W$-$t$-$b^\prime$ coupling ($|V_{tb^\prime}|^2$ in the
model with a fourth family) and the mass of the $b^\prime$.  In 
Figure~\ref{bprimeratefig} we present the NLO rate for $t \bar{b^\prime}$
production (as well as $\bar{t} b^\prime$ production) without any
decay BR's.
Since the
$|V_{tb^\prime}|^2$ dependence may be factored out, these rates
assume $V_{tb^\prime} = 1$.  
The collider signatures resulting from
such a process depend on the decay modes available to the $b^\prime$.
If $m_{b^\prime} > m_t + m_W$, it is likely to decay into a top quark and
a $W^-$, and the events will have a $t \, \bar{t}$ pair with an additional
$W^\pm$ boson.  If this decay mode is not open, loop induced decays
such as $b^\prime \ra b \, \gamma$ may become important, resulting in
a signature $t \, \bar{b}$ plus a hard photon whose invariant mass
with the $b$ quark will reconstruct the mass of the $b^\prime$.

\subsection{Extra Gauge Bosons}
\indent \indent

Another simple extension of the SM is to postulate the existence of
a larger gauge group which somehow reduces to the SM gauge
group at low energies.  Such theories naturally have additional 
gauge bosons, some of which may prefer to couple to the top (or even
the entire third family).  Examples of such theories include 
the top-color \cite{topcolor} and top-flavor 
\cite{topflavor,tf1,tf2} models, and gauged-flavor symmetry models
\cite{flavoron}
which give special dynamics to the third family in order to explain the
large top mass.  As a specific example, we will consider the 
top-flavor
model with an extra ${\rm SU(2)}_h$ gauge symmetry that generates
a top mass through a seesaw effect \cite{tf2}.

This model has an
over-all gauge symmetry of SU(3$)_{C} \times$
SU(2$)_{h} \times$ SU(2$)_{l} \times$ U(1$)_{Y}$, and thus
there are three additional weak
bosons ($W^{\prime \pm}$ and $Z^{\prime}$).
The first and second generation fermions 
and third family leptons transform under
SU(2$)_{l}$, while the third generation quarks transform
under SU(2$)_{h}$.  As was alluded to before,
in order to cancel the anomaly and provide
a seesaw mechanism to generate the top mass, an additional
doublet of heavy quarks whose left-handed components transform
under ${\rm SU(2)}_l$ and right-handed components transform
under ${\rm SU(2)}_h$ is also present.

A set of scalar fields transforming under
both SU(2$)_{l}$ and SU(2$)_{h}$ acquire a VEV, $u$,
and break the symmetry to SU(2$)_{l + h} \times$
U(1$)_{Y}$.  From here the usual electro-weak symmetry breaking
can be accomplished by introducing a scalar doublet which acquires
a VEV $v$, further breaking the gauge symmetry to U(1$)_{EM}$.
We write the
covariant derivatives for the fermions as,
\bea
D^{\mu} &=& \partial^{\mu} +
i g_{l} \; T^{a}_{l} \; {W^{a}}^{\mu}_{l} +
i g_{h} \; T^{a}_{h} \; {W^{a}}^{\mu}_{h} +
i g_1 \; \frac{Y}{2} \; B^{\mu} ,
\eea
where $T^{a}_{l (h)}$ are the generators for SU(2$)_{l (h)}$,
$Y$ is the hyper-charge generator, and ${W^{a}}^{\mu}_{l (h)}$
and $B^{\mu}$ are the gauge bosons for the SU(2$)_{l (h)}$ and
U(1$)_{Y}$ symmetries.  The gauge couplings may be written,
\bea
g_{l} = \frac{e}{\sin \theta_W \cos \phi} \, , \hspace{1cm}
g_{h} = \frac{e}{\sin \theta_W \sin \phi} \, , \hspace{1cm}
g_1 = \frac{e}{\cos \theta_W} \, ,
\eea
where $\phi$ is a new
parameter in the theory.  Thus this theory is determined
by two additional quantities $x = u / v$,
the ratio of the two VEV's, and $\sin^2 \phi$, which
characterizes the mixing between the heavy and light SU(2)
gauge couplings.

At leading order, the heavy bosons
are degenerate in mass,
\bea
{M^2}_{Z^{\prime}, W^{\prime}} =
{M_0}^2 \left( \frac{x}{\sin^2 \phi \cos^2 \phi}
+ \frac{\sin^2 \phi}{\cos^2 \phi} \right) ,
\eea
where 
${M_0}^2 = \frac{ e^2 v^2 }{4 \sin^2 \theta_W \cos^2 \theta_W}$.
We can thus parameterize the model by the
heavy boson mass, $M_{Z^{\prime}}$, and the
mixing parameter\footnote{As shown in \cite{topflavor},
for $\sin^2 \phi \leq$ 0.04, the
third family fermion coupling to the heavy gauge bosons can
become non-perturbative.  Thus we restrict ourselves to considering
$0.95 \geq \sin^2 \phi \geq$ 0.05.},
$\sin^2 \phi$.
Low energy data requires that the mass
of these heavy bosons, $M_{Z^{\prime}}$,
be greater than about $900$ GeV \cite{tf1}.

\begin{figure}[t]
\epsfysize=1.5in
\centerline{\epsfbox{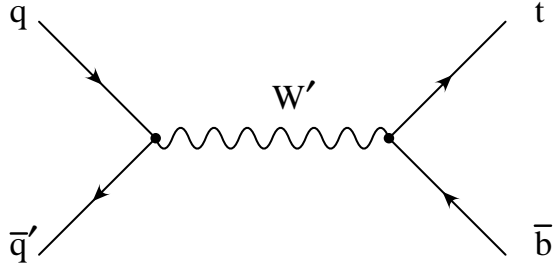}}
\caption{Feynman diagrams illustrating how a $W^\prime$ boson
can contribute to single top production through
$q \, \bar{q}^\prime \ra W^\prime \ra t \, \bar{b}$.}
\label{wpfig}
\end{figure}

\begin{figure}[p]
\epsfysize=5.0in
\centerline{\epsfbox{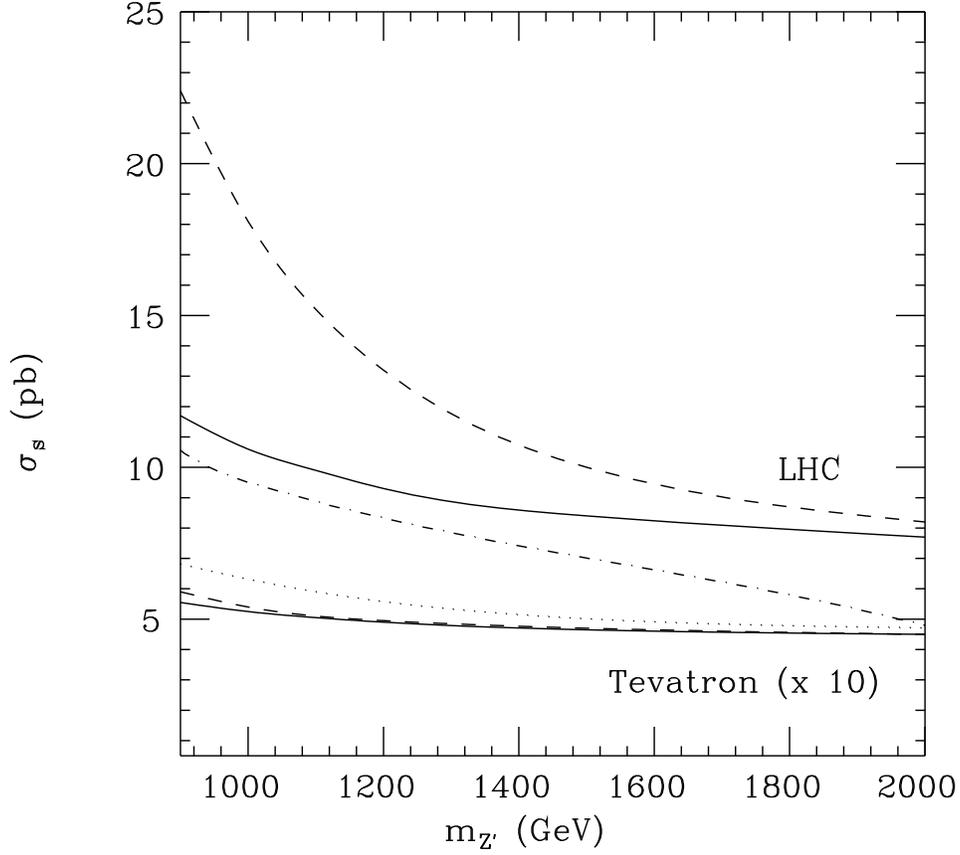}}
\caption{The NLO rate of 
$q \, \bar{q}^\prime \ra W, W^\prime \ra t \, \bar{b}$
($\sigma_S$) in pb
at the Tevatron (lower curves)
and LHC (upper curves), for the top-flavor model with
$\sin^2 \phi = 0.05$ (solid curves) and 
$\sin^2 \phi = 0.25$ (dashed curves), as a function of
$M_{Z^\prime} = M_{W^\prime}$.
The Tevatron cross sections are multiplied by a factor of 10.
At the Tevatron, the $\bar{t}$ production rate is equal to the
$t$ rate.  At the LHC the $\bar{t}$ rates are shown for
$\sin^2 \phi = 0.05$ (dotted curve) and
$\sin^2 \phi = 0.25$ (dot-dashed curve).}
\label{wpratefig}
\end{figure}

The additional $W^\prime$ boson can contribute to the $s$-channel mode
of single top production through virtual exchange of a $W^\prime$
as shown in Figure~\ref{wpfig}
\cite{es}.  In particular, if enough energy is
available, the $W^\prime$ may be produced close to on-shell, and a
resonant enhancement of the signal may result.
Since the additional
diagrams involve a virtual $W^\prime$, they will
interfere with the SM $W$-exchange diagrams, and thus the net
rate of single top
production can be increased or decreased as a result, though the
particular model under study always results in an increased $s$-channel
single top rate\footnote{One interesting example of a model with a
$W^\prime$ that interferes non-trivially with the SM $W$ exchange in single
top production is provided by embedding the gauge fields in extra dimensions
\cite{tevgut,singletxd}.}. In Figure~\ref{wpratefig} the resulting 
NLO $s$-channel rate
for $q \, \bar{q}^\prime \ra W, W^\prime \ra t \, \bar{b}$ at Tevatron
and LHC is shown, as a function of the $W^\prime$ mass, 
for a few values of $\sin^2 \phi$.  The rate for $\bar{t}$ production
through the same process is shown as well.
While the final state particles for this case
are the same as the SM $s$-channel mode, the distribution of the invariant
mass of the $t \, \bar{b}$ system could show a Breit-Wigner
resonance effect around $M_{W^\prime}$, which serves to identify this type
of new physics.  However, if the mass of the $W^\prime$ is large compared
to the collider energy, and its width broad, the resonance shape can
be washed out even at the parton level.
Jet energy smearing from detector
resolution effects will further degrade the resonance 
and could make it difficult to identify.

A $t$-channel
exchange of the $W^\prime$ is also possible, but in that case a negligible
effect is expected because the boson must have a space-like momentum,
and thus the additional contributions are suppressed by $1 / M^2_{W^\prime}$,
and are not likely to be visible.  This argument applies quite
generally to any heavy particle's effect on single top production.  The 
$s$-channel rate is quite sensitive to a heavy particle because of the
possibility of resonant production, whereas the $t$-channel rate
is insensitive because the space-like exchange is suppressed by the
heavy particle mass.

Clearly, the existence of a $W^\prime$ will not influence the rate of
$t \, W^-$ production, but it could allow for exotic production modes
such as $b \, g \ra t \, {W^\prime}$.  If the ${W^\prime}$ has a strong
coupling with the third family, then one would expect that its dominant
decay should be into $b \, \bar{t}$, and thus a final state of
$t \, \bar{t} \, b$ would result with the $t \, \bar{b}$ invariant mass
reconstructing the $W^\prime$ mass.
Current limits on the $W^\prime$ mass in the
top-flavor model make this mode non viable at the Tevatron
and unpromising at the LHC,
with a cross section of 1.14 pb
for $M_{W^\prime} = 900$ GeV
and $\sin^2 \phi = 0.05$ including the large QCD logarithmic corrections
described in \cite{mytw}. 
However, an observation of this signal would be a clear indication
of the nature of the new physics.  

\subsection{Extra Scalar Bosons}
\indent \indent

Scalar particles appear in many theories,
usually associated with the spontaneous breaking of a symmetry
in a Lorentz invariant fashion.
In the SM and the minimal
supersymmetric extension, fundamental scalar fields of both neutral
and charged character are present in the theory, and 
some are expected to
have a strong coupling with the top because of the role they play in
generating fermion masses.  In dynamical models such as the top-condensate
and top-color assisted technicolor models, scalar particles exist as
bound states of top and bottom quarks 
\cite{topcondensate,topcolor,topflavor,tf1,tf2}.
These composite scalars also
have a strong coupling to the top because of their role in the generation
of the top mass.  
Another way to look at this is that the top mass is large because the
strong forces needed to bind tops into Higgs result in a strong
Higgs coupling to top.
This illustrates the fact that the large top mass
naturally makes it a likely place to look for physics associated with the
EWSB.

\begin{figure}[t]
\epsfysize=1.5in
\centerline{\epsfbox{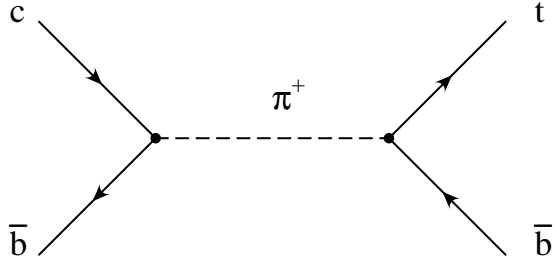}}
\caption{Feynman diagram illustrating how a charged top-pion
can contribute to single top production through 
$c \, \bar{b} \ra \pi^+ \ra t \, \bar{b}$.}
\label{toppionfig}
\end{figure}

An example is provided by the charged composite
top-pions ($\pi^\pm$)
of the top-color model,
which can be produced in the $s$-channel through
$c \, \bar{b}$ fusion \cite{toppion}, 
$c \, \bar{b} \ra \pi^+ \ra t \, \bar{b}$.
The leading order Feynman diagram is shown in Figure~\ref{toppionfig}.
In this case the strong $\pi^+$-$c$-$\bar{b}$ coupling comes 
from mixing
between the right-handed $t$ and $c$ quarks.  
The CKM matrix is 
the product of the left-handed rotation matrices 
for the up- and down-type quarks, so it does not constrain a possibly
large right-handed $t$ and $c$ mixing.
The fact that 
this interaction has a right-handed nature will prove
interesting when we study top polarization below.

\begin{figure}[p]
\epsfysize=5.0in
\centerline{\epsfbox{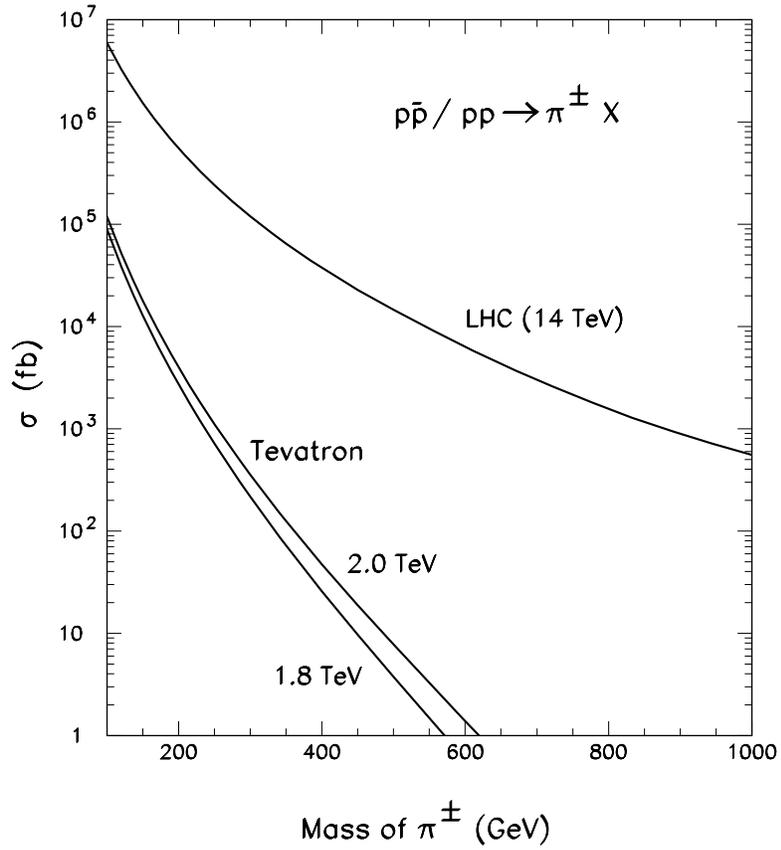}}
\caption{The NLO rate of single top production through the
reaction $c \, \bar{b} \ra \pi^+ \ra t \, \bar{b}$
as a function of $M_{\pi^\pm}$, assuming a $t_R$-$c_R$ mixing
of $20\%$.  These rates include $t$ and $\bar{t}$
production, which are equal for both Tevatron and LHC.}
\label{toppionratefig}
\end{figure}

Like the $W^\prime$, the $\pi^\pm$ contributes to the $s$-channel topology
of single top production and can allow large resonant contributions.
However, unlike the $W^\prime$, the $\pi^+$ does not interfere
with the SM amplitudes ($q' {\bar q} \ra t {\bar b}$), 
because the SM contribution arises dominantly from (left-handed) 
light quarks.
  In Figure~\ref{toppionratefig}, we
present the NLO single top rate from the top-pion process \cite{nlotoppion},
for a variety of $\pi^\pm$ masses with the $t_R$-$c_R$ mixing set equal to
$20\%$.  
The two other modes of single top production are once again relatively
insensitive to the $\pi^\pm$.  The $t$-channel process is insensitive 
because its contribution is
 suppressed by $1/ M^2_{\pi^\pm}$ and the fact that the $\pi^\pm$
does not couple to light quarks.  The $t \, W^-$ mode is insensitive because
presumably the $\pi^\pm$ is generally distinguishable from a 
$W^\pm$ boson. For example, $g \, b \ra \pi^- \, t \ra \bar{t} \, b \, t$
will not be mistaken for $t \, W^-$ production when $\pi^-$ 
predominantly decays into $\bar{t}$ 
and $b$, as in the topcolor model.

Single top quarks can also be produced by neutral top-pions 
(produced for example from $g g \ra \pi^0$ decaying into
$t \bar{c}$.  Again, one expects to see a large
effect in the $s$-channel single top mode, and no effect in the
$t$-channel and \tw modes, because of the large $\pi^0$ mass and 
small coupling to light quarks \cite{toppi0}.

\begin{figure}[t]
\epsfysize=1.5in
\centerline{\epsfbox{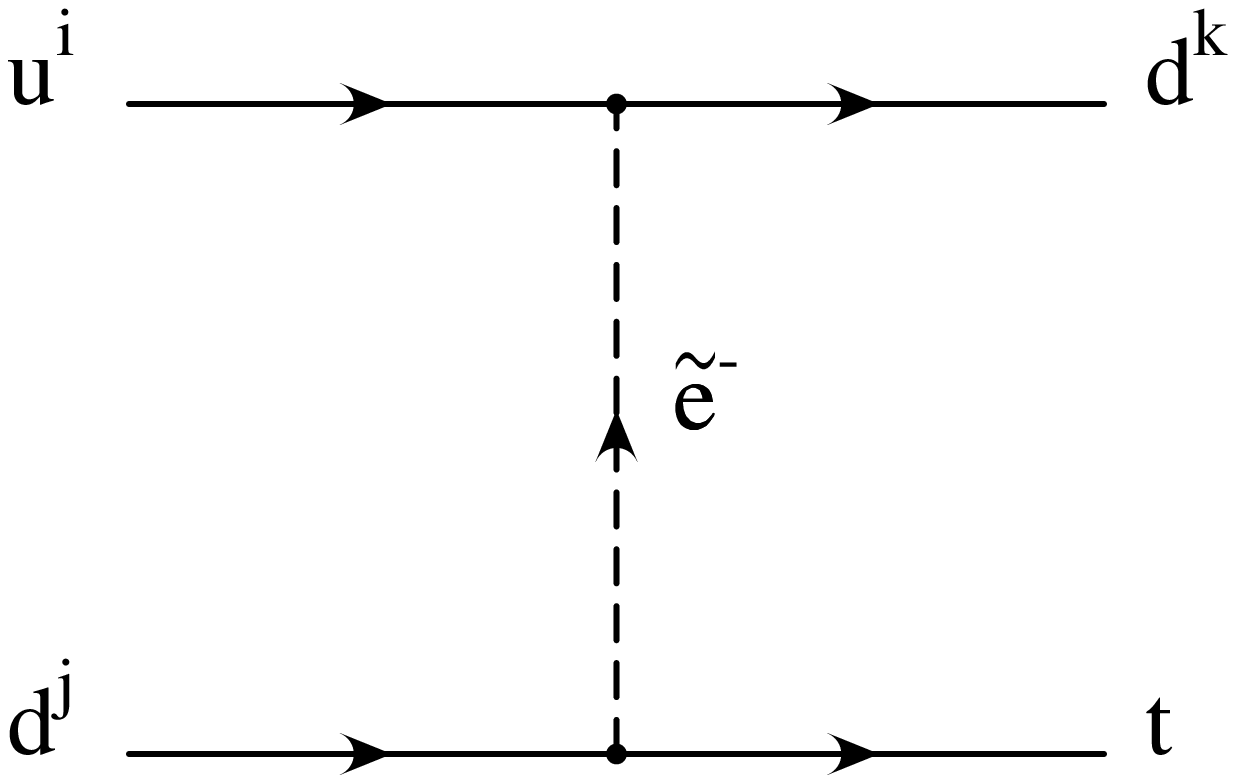}}
\caption{Illustrative Feynman diagram for the $R$-parity violating
production of a single top quark through the reaction 
$u^i \, d^j \ra t \, d^k$.}
\label{rpviolfig1}
\end{figure}

\begin{figure}[t]
\epsfysize=1.5in
\centerline{\epsfbox{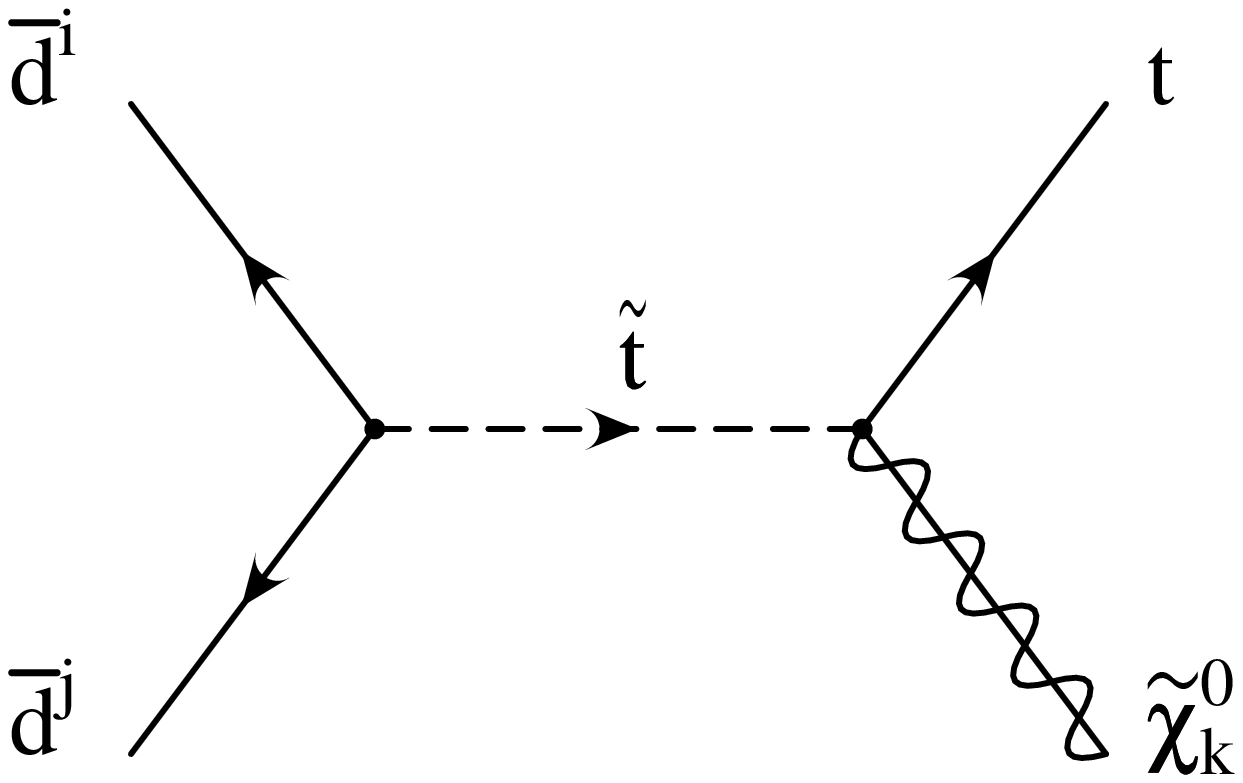}}
\caption{Illustrative Feynman diagram for the $R$-parity violating
production of a single top quark through the reaction 
$\overline{d}^i \, \overline{d}^j \ra \widetilde{t} \ra t \, 
\widetilde{\chi}^0_k$.}
\label{rpviolfig2}
\end{figure}

Different types of scalar particles that couple top and bottom can
be analyzed in a similar fashion.  The $s$-channel mode allows for
resonant production, which can show a large effect, whereas the
$t$-channel mode is suppressed by the space-like momentum 
(and large mass) of the
exchanged massive particle.  The $t \, W^-$ mode is insensitive because
in that case the $W$ is actually observed in the final state.
One example of this kind are the technipions in a technicolor model 
which can contribute to single top production in this way \cite{tcst}.  
Another example is provided by SUSY models
with broken $R$-parity, in which
the scalar partners of the leptons (sleptons)
can couple with the top and bottom quarks, and will contribute to
single top production \cite{rpviol}, or through reactions such as
$\bar{d} \, \bar{s} \ra \widetilde{t} \ra t \, \widetilde{\chi}^0_i$
\cite{edbrizack}. (See Figures~\ref{rpviolfig1} and 
\ref{rpviolfig2} for their representative Feynman
diagrams.)

\begin{figure}[t]
\epsfysize=1.5in
\centerline{\epsfbox{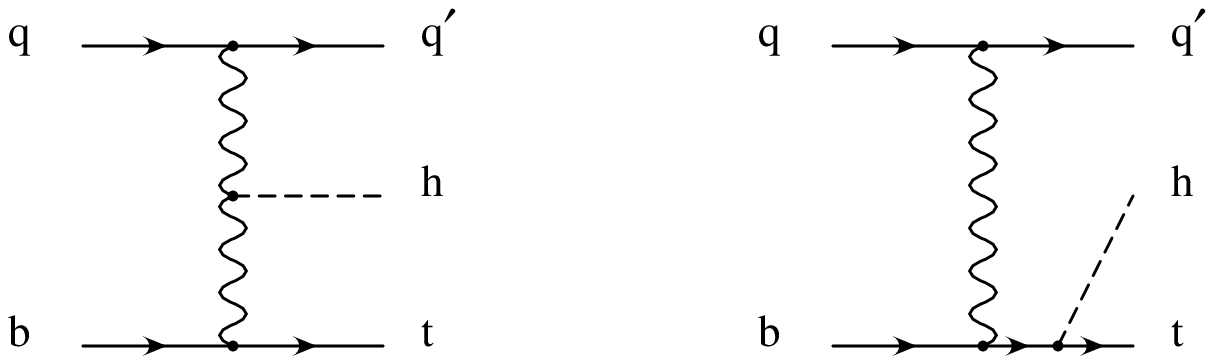}}
\caption{Feynman diagrams for associated production of a
neutral scalar and single top quark:
$q \, b \ra q^\prime \, t \, h$.}
\label{hstfig}
\end{figure}

\begin{figure}[t]
\epsfysize=4.0in
\centerline{\epsfbox{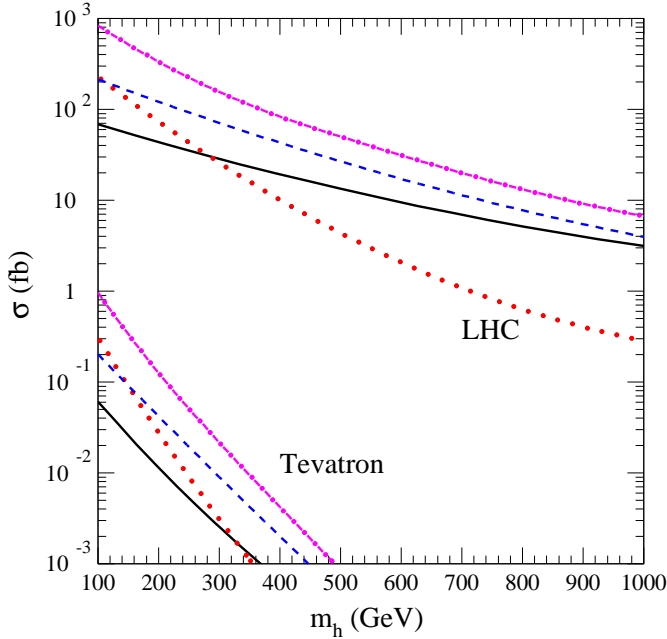}}
\caption{The cross section for production of Higgs in association with a 
single top quark, as a function of the Higgs mass at the Tevatron 
Run~II (lower family of curves) and the LHC (upper family of curves).
The cross sections are shown for models in which the Higgs couplings
are completely SM-like (solid curves), the coupling to
the $W^\pm$ boson is zero and the coupling to top is SM-like (dotted curves),
the coupling to top is zero and the coupling to $W^\pm$ is SM-like (dashed
curves), and the coupling to $W^\pm$ is SM-like and the coupling to top
is $-1$ times the SM coupling (dash-dotted curves).}
\label{sthratefig}
\end{figure}

As a final note, there is the interesting process in which a neutral
scalar (like the Higgs boson of the SM) is produced in association with a
single top quark \cite{hsinglet}.
Feynman diagrams are shown in Figure~\ref{hstfig}.
This process is of interest because while the magnitude of the $h$-$W$-$W$
and $h$-$t$-$t$ couplings can be measured independently by studying 
$q \, \bar{q}^\prime \ra W^* \ra W \, h$ and 
$q \, \bar{q} \: (g \, g) \ra h \, t \, \bar{t}$
(or the equivalent processes at a high energy lepton collider), 
the relative phase between
the couplings can be found from the process $q \, b \ra q^\prime \, t \, h$,
as that phase information is contained in the interference between the
two diagrams shown in Figure~\ref{hstfig}.  This process is extremely small
compared to the other two mentioned (with a SM cross section of 
$4.4 \times 10^{-2}$ fb at the
Tevatron and 0.06 pb at the LHC for a Higgs boson of mass
($m_h$) 110 GeV, and
including both $t$ and $\bar{t}$ production),
and thus it is not promising a discovery
mode.  The small SM rate results from the large destructive
interference between the two diagrams.  Typically, it
yields a reduction
in the rate as compared to the contribution from each individual diagram
by about an order of magnitude.
In Figure~\ref{sthratefig} we show the contributions from 
each Feynman diagram
containing either $h$-$t$-$t$ or $h$-$W$-$W$ vertex to the total
production rate of $q' h t$ as a function of $m_h$ at Run II of 
the Tevatron and the LHC.
Though the rates at the Tevatron are very small in the SM, with the
enhanced coupling of $h$-$t$-$\bar{t}$ 
predicted by some models of new physics, this process could conceivably
be observed there.
The strong cancellation predicted by the SM  
indicates that this process is very 
sensitive to any physics that modifies the relative phase (and size)
between the $h$-$t$-$t$ and $h$-$W$-$W$ couplings from the SM relation.
Thus, it contains important information not available in the 
$W h$ and $h t \bar t$ processes, and should be carefully tested
experimentally.

\section{Modified Top Quark Interactions}
\label{modint}
\indent \indent

Another interesting set of properties of the top that can be 
studied in single top production are the top couplings to light particles.
The electroweak chiral Lagrangian \cite{ewcl} (EWCL) provides
a powerful way to study such effects model-independently.
Following the EWCL approach, we write an effective Lagrangian to describe
low energy physics as,
\bea
 {\cal L}_{eff} &=& {\cal L}_{SM} + {\cal L}_4 + {\cal L}_5 + ... ,
\eea
where ${\cal L}_{SM}$ refers to the usual SM Lagrangian, 
and ${\cal L}_4$ and ${\cal L}_5$ are the Lagrangians
containing deviations from the SM in terms of operators of
mass dimension 4 and 5, respectively.  The spirit of the EWCL
approach is that higher order operators will generally be suppressed
by higher powers of $\Lambda$, the scale at which the effective
theory break down.  Thus, for low energy processes 
occuring at energies below $\Lambda$ the lowest dimension
anomalous operator is expected to provide the largest effect
\cite{nda}.
We choose the effective Lagrangian to realize the weak symmetry
nonlinearly, as this is the most general possibility 
\cite{nonlinear}.  This is
most appropriate for a strongly interacting underlying theory,
which may not contain a Higgs doublet in any real sense.  If
the underlying theory is weakly coupled, it is most likely
more appropriate to describe the new physics effects in terms of
a theory with the weak symmetry realized linearly, in which case
all nonstandard effects will be suppressed by at least one power
of $\Lambda$.

Terms which
have the potential to modify single top production include
mass dimension 4 operators \cite{ehabewcl},
\bea
  \label{ewcleq1}
  {\cal L}_4 &=& \frac{ e }
  {\sqrt{2} \, \sin \, \theta_W} 
  {W^+}_\mu \left( 
    \kappa_{Wtb}^L \, e^{i \, \phi_{Wtb}^L} \,
    \bar{b} \, \gamma^\mu \, P_L \, t +
    \kappa_{Wtb}^R \, e^{i \, \phi_{Wtb}^R} \,
    \bar{b} \, \gamma^\mu \, P_R \, t \right) \\[0.3cm] & & \,
    + \frac{e}
    {2 \, \sin \theta_W \, \cos \theta_W} 
    {Z}_\mu \left( \;
    e^{i \, \phi_{Ztc}^L} \,
    \kappa_{Ztc}^L \, \bar{c} \, \gamma^\mu \, P_L \, t \:+
    e^{i \, \phi_{Ztc}^R} \,
    \kappa_{Ztc}^R \, \bar{c} \, \gamma^\mu \, P_R \, t \; 
    \right) + H.c. , \nonumber
\eea
which can be classified as two charged current
operators which modify the SM top weak interactions with the $W$ boson and
$b$ quark, 
as well as two flavor-changing neutral current (FCNC) operators
involving the $Z$ boson, $t$, and $c$ quarks.
Additional dimension 4 FCNC operators with the $c$ quark replaced by the
$u$ quark are also possible.
We have included the CP violating
phases $\phi_{Wtb(Ztc)}^{L(R)}$ in the interactions
for generality, though they are not always considered in the literature.

We note that the anomalous $Z$-$t$-$c$ and $W$-$t$-$b$ interactions 
would have first appeared at
dimension 6 if the electroweak symmetry were realized linearly.  This
would lead to the estimates 
$\kappa_{Wtb}, \kappa_{Ztc} \sim {\cal O}(v^2/\Lambda^2)$ where $v$ is
the electroweak symmetry breaking VEV.  There are also operators such
as $(\bar t \, c) \, (\bar c u)$, a four fermion contact interaction,
which first appear at dimension 6,
the linearly realized situation should also include these possibilities
in the analysis.  These genuine dimension 6 operators have very different
energy dependence than the dimension 4 operators we have considered,
which might allow one disentangle their contributions by careful study
of the kinematics of the single top events.

In addition there are dimension 5 operators that involve interactions
between new sets of particles and the top\footnote{There are also dimension
five operators involving the sets of particles that already appear in
Equation~(\ref{ewcleq1}) \cite{dim5}.  As discussed above,
naive dimensional analysis suggests that
these operators are less significant than their
dimension four counterparts, so we limit ${\cal L}_5$ to the dimension
5 operators which involve only new sets of fields.}
and can contribute to single top production.  These include
the FCNC operators,
\bea
  \label{ewcleq2}
  {\cal L}_5 &=& \frac{ g_S \, G^a_{\mu \nu} }
    { \Lambda_{gtc} }  \left( \:
    e^{i \, \phi_{gtc}^L} \;
    \kappa_{gtc}^L \: \bar{c} \: T^a \;
    \sigma^{\mu \nu} \, P_L \, t
  + \; e^{i \, \phi_{gtc}^R} \;
    \kappa_{gtc}^R \: \bar{c} \: T^a \; 
    \sigma^{\mu \nu}
    \, P_R \, t
    \right) 
    \\[0.3cm] & & \,
  + \frac{2 \, e \, F_{\mu \nu} }{3 \, \Lambda_{\gamma tc} }
    \left( \; e^{i \, \phi_{\gamma tc}^L} \;
    \kappa_{\gamma tc}^L \, \bar{c} \:
    \sigma^{\mu \nu} \, P_L \, t
  + e^{i \, \phi_{\gamma tc}^R} \;
    \kappa_{\gamma tc}^R \, \bar{c} \:
    \sigma^{\mu \nu}
    \, P_R \, t
    \right) \nonumber + H.c.,
\eea
which couple the charm quark to the top and gluon or photon
fields.  Once again, we have included $CP$ violating phases
$\phi_{gtc (\gamma tc)}^{L(R)}$ which are not generally considered
in the literature.
Additional operators with the charm replaced by the up
quark are also possible.
As dimension 5 operators, these terms have couplings with
dimension of inverse mass that have been written in the form of
$\kappa_{gtc}^A / \Lambda_{gtc}$ and 
$\kappa_{\gamma tc}^A / \Lambda_{\gamma tc}$, where $A = L,R$.
In the discussion below, we will consider only the cases where
all of these $\kappa$ couplings are 1, and consider the magnitude
allowed for the scales $\Lambda_{gtc}$ and $\Lambda_{\gamma tc}$.
If the underlying theory is strongly coupled, these mass scales
may be thought of as the energy scale in which the SM breaks down
and must be replaced with the underlying theory.  However, 
it should be kept in mind that if the
underlying theory is weakly coupled, this interpretation is somewhat
obscured by the fact that the energy scales $\Lambda$ will also include
small factors of the fundamental 
(weak) interaction strength and loop suppression
factors.  Even in this case, an experimental constraint on $\Lambda$
is very useful because it will provide constraints on the parameters
of an underlying model of new physics.

The dimension 4 terms which modify the $W$-$t$-$b$ vertex will clearly
have a large impact on single top production \cite{dthesis}.  However,
$\kappa_{Wtb}^R$ is already very strongly constrained 
by low energy
$b \ra s \, \gamma$ data \cite{cleo}, which requires \cite{kappaR},
\bea
  -.0035 \geq ( \; \kappa_{Wtb}^R \cos \phi_{Wtb}^R
  + 20 \, {\kappa_{Wtb}^R}^2 \; ) \leq 0.0039 ,
\eea
provided that $\kappa_{Wtb}^L$ is smaller than 0.2.
Further, the CP-violating observable that measures the asymmetry in the
decay rates of $b \rightarrow s X$ and ${\bar b} \rightarrow {\bar s} X$
can test the imaginary part of the right-handed charged current
coupling, i.e. $\kappa_{Wtb}^R \sin \phi_{Wtb}^R$ \cite{kappaR}.
Given this strong constraint, it is
unlikely that further information about $\kappa_{Wtb}^R$
can be gleaned from single top production, so
we will assume $\kappa_{Wtb}^R = 0$ in the discussion below.
Although the $b\to s \gamma$ process does not provide a
good test of the left handed, CP-odd, $W$-$t$-$b$ coupling,
i.e. $\kappa_{Wtb}^L \sin \phi_{Wtb}^L$, 
there are other $B$-decay processes with a good potential to measure 
it at future $B$ factories.  In particular, the hadronic channels
$B_d \ra \phi K_s$ and $B_d \ra \Psi K_s$ have been considered
in Ref.~\cite{valencia}.
Assuming $\kappa_{Wtb}^R=0$, 
all three single-top production modes are sensitive to $\kappa_{Wtb}^L$,
and will be proportional to 
$(1 + {\kappa_{Wtb}^L}^2 \cos^2 \phi^L_{Wtb} 
+ 2 \, \kappa_{Wtb}^L \cos \phi^L_{Wtb})$ 
much the same way that
they will all be sensitive to $V_{tb}$ in the SM\footnote{This is because 
the dimension 4 term that is proportional to $\kappa_{Wtb}^L$ in 
${\cal L}_4$ does not depend on the momenta of the interacting particles,
as is the case for the SM $W$-$t$-$b$ interaction.  
For higher dimension $W$-$t$-$b$ operators,
which may depend on the momenta, each single top mode will respond
differently to the new interaction, and thus could be used to distinguish
one operator from another.}.

\begin{figure}[t]
\epsfxsize=6.0in
\centerline{\epsfbox{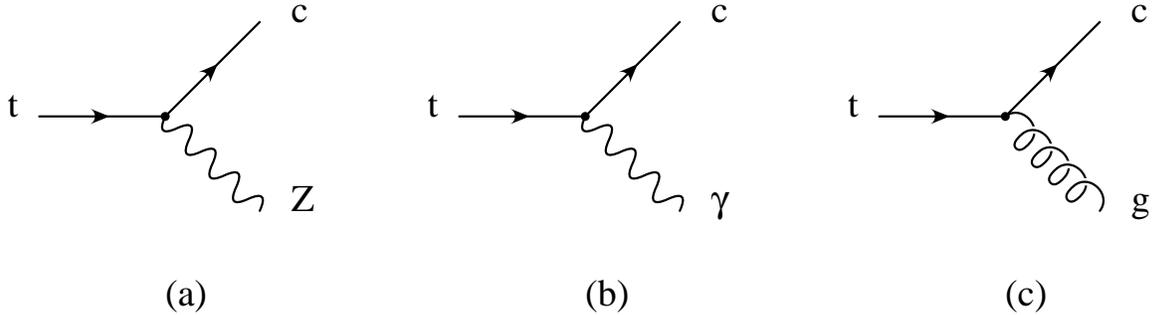}}
\caption{Feynman diagrams showing FCNC top decays through
(a)~$t \ra Z \, c$, (b)~$t \ra \gamma \, c$, and (c)~$t \ra g \, c$.}
\label{tncdecayfig}
\end{figure}

The flavor-changing neutral current terms in ${\cal L}_4$ and ${\cal L}_5$
will also contribute to single top production, and since they involve particles
lighter than the top mass, will also contribute to top decays
through Feynman diagrams such as those shown in Figure~\ref{tncdecayfig},
which illustrate FCNC $t$ decays to $c$.  The FCNC interactions between
$t$ and $u$ will allow for exotic decays of the same type, but with
the $c$ quark exchanged with a $u$ quark.
One
could hope to learn about these anomalous FCNC couplings both by studying
single top production and top decays.  However, this brings us back to the
problem with using top decays to determine the magnitude of a coupling -
the decay can provide information about the relative branching fraction of
the exotic decay compared to the SM top decay $t \ra W^+ \, b$, but since
it does not allow one to measure the top decay width, it cannot provide
a limit on the size of the exotic operator without 
first making an assumption
concerning the nature of the $W$-$t$-$b$ interaction.
In fact, one might think that single top would suffer from the same
difficulty in distinguishing the magnitude of
new physics in the $W$-$t$-$b$ interaction from new physics in a
FCNC interaction.  However, as we shall see, one can use the three
modes of single top production separately to disentangle the FCNC
new physics from the possibility of $W$-$t$-$b$ new physics.

The three FCNC operators have a similar structure of a light $c$ (or $u$)
quark interacting with a top and a neutral vector boson.  Thus, we can
discuss their impact on the three single top processes rather generally
by considering the specific example of the $Z$-$t$-$c$ operator.
In examining the FCNC operators in Equations~(\ref{ewcleq1}) and
(\ref{ewcleq2}), we note that they can have left-handed and
right-handed interactions with different interaction coefficients
(and even different phases).
For now we will restrict our discussion to the case where
all of the phases are zero, and discuss only the magnitude of the
interactions, set by $\Lambda_{gtc}$, $\Lambda_{\gamma tc}$, 
and $\kappa_{Ztc}$.  We will return to the subject of exploring their
chiral structure when we consider top polarization in 
Section~\ref{polarization}.

\begin{figure}[t]
\epsfysize=1.4in
\centerline{\epsfbox{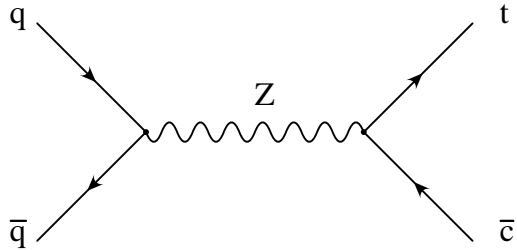}}
\caption{Feynman diagram showing how a FCNC $Z$-$t$-$c$ interaction
contributes to the $s$-channel mode of single top production
through $q \, \bar{q} \ra Z^* \ra t \, \bar{c}$.}
\label{schanztcfig}
\end{figure}

The $Z$-$t$-$c$ 
operator will allow for additional contribution to the $s$-channel
mode of single top production through reactions such as 
$q \, \bar{q} \ra Z^* \ra t \, \bar{c}$,
shown in Figure~\ref{schanztcfig}.  This reaction has different
initial and final state from the SM $s$-channel mode, and thus 
this new physics contribution does not interfere with the SM contribution
to single top production.
(The FCNC $Z$-$t$-$c$ coupling can be induced at the loop level in the
SM via the CKM mechanism, but its magnitude is small and can be
neglected in our analysis.)
The fact that the new physics process has a $\bar{c}$ instead of a
$\bar{b}$ in the final state has a drastic practical consequence that the
new physics production mechanism probably cannot be experimentally extracted
at all, because in order to separate the $s$-channel mode from the large
$t \, \bar{t}$ and $W$-gluon fusion backgrounds, it is necessary to tag
the $\bar{b}$ produced in association with the top in the $s$-channel
mode, in addition to the $b$ from the top decay.  Thus, while a FCNC operator
could contribute to $s$-channel production of a single top, it will not be
counted as such\footnote{It could be possible to search for $s$-channel
production via a FCNC with a specialized strategy differing from the
usual one employed to extract the $W^*$ process, but such a search
will require identifying the ${\bar c}$ produced in association with the
top, and will suffer from large backgrounds from $t \, \bar{t}$ and
$W$-gluon fusion single top processes.}.

\begin{figure}[t]
\epsfysize=1.8in
\centerline{\epsfbox{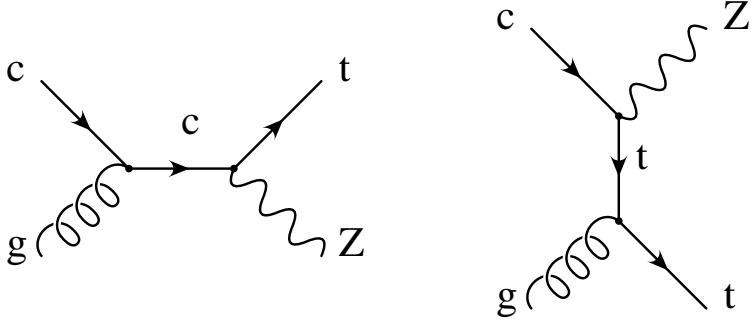}}
\caption{Feynman diagrams showing how a FCNC $Z$-$t$-$c$ interaction
contributes to the exotic mode of single top production
$g \, c \ra t \, Z$.}
\label{gctZfig}
\end{figure}

The $t \, W^-$ mode cannot receive a contribution from a FCNC, though
a FCNC will generally allow for new exotic production mechanisms such
as $g \, c \ra t \, Z$ shown in Figure~\ref{gctZfig}
\cite{ztprod}.  From this
consideration, along with the analysis of the $t \, W^-$ mode in
Section~\ref{extrap}, we see that the $t \, W^-$ mode has a special
quality because both the top and the $W$ are in the final state 
(and thus identifiable).  Thus, it is sensitive to new physics which
modifies the $W$-$t$-$b$ interaction\footnote{
Of course it is also sensitive to the $W$-$t$-$s$ and $W$-$t$-$d$
interactions.}, but it is not sensitive to
nonstandard physics involving new particles or FCNC's.  Thus, the
$t \, W^-$ mode represents a chance to study the $W$-$t$-$b$
vertex without contamination from FCNC new physics.

\begin{figure}[t]
\epsfysize=1.8in
\centerline{\epsfbox{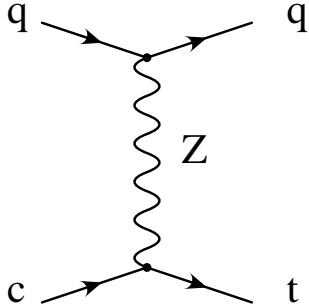}}
\caption{Feynman diagram showing how a FCNC $Z$-$t$-$c$ interaction
contributes to the $t$-channel mode of single top production
through $c \, q \ra t \, q$.}
\label{tchanztcfig}
\end{figure}

The $W$-gluon fusion mode of single top production is quite sensitive to
a FCNC involving the top and one of the light partons, through
processes such as $c \, q \ra t \, q$, from Feynman diagrams
such as those shown in Figure~\ref{tchanztcfig}.
The FCNC operators involve a different set of spectator quarks
in the reaction, and thus they do not interfere with the SM
$t$-channel process.  In fact,
because the $W$-gluon fusion mode
requires finding a $b$ inside a hadron, which has less probability
than finding a lighter parton, the FCNC's involving $u$ or
$c$ quarks already receive an enhancement relative to the SM $t$-channel
rate purely from the larger parton densities
for the lighter flavors.  This can somewhat compensate
for a (presumably) smaller FCNC coupling.  This shows the sense in which
the $t$-channel single top mode is sensitive to the top quark's decay
properties.  The same type of new physics which opens up new top decay
modes (and thus modifies the top's total width) will also modify the
$t$-channel rate of single top production, because the same light partons
into which the top may decay are also responsible for producing
single tops in the $t$-channel process.  Thus, one can think of
the $t$-channel process as a kind of measure of the 
inclusive top width.

Because of the strong motivation to use single top production to
study FCNC operators involving the top quark, detailed
simulations of the effect of the $g$-$t$-$c$ operator on single top
production were performed \cite{tcgmenehab}, and found that this operator
could be constrained by the process $q \, \bar{q} \ra t \, \bar{c}$ to
$\Lambda_{gtc} \geq 4.5$ 
TeV at Run~II of the Tevatron if no new physics
signal were to be found.  Further refinements on this idea
\cite{tcgother} showed that it could be improved by including other
reactions such as $g \, c \ra t$, $g \, c \ra g \, t$,
$q \, c \ra t \, q$, and $ g \, g \ra t \, \bar{c}$
to $\Lambda_{gtc} \geq 10.9$ TeV at the Tevatron Run~II
and to $\Lambda_{gtc} \geq 164$ TeV at the LHC.

Detailed simulations of the $Z$-$t$-$c$ and $\gamma$-$t$-$c$ operators
have so far been confined to studies of top decays
\cite{tcgdecay,tcz,tcadecay}.
The left-handed $Z$-$t$-$c$ FCNC operator, whose strength is
parameterized in Equation (\ref{ewcleq1}) by the
quantity $|\kappa_{Ztc}^L|$ is constrained by
low energy data on flavor-mixing processes
to be less than the order of magnitude of 0.05 \cite{tcz}.
The right-handed operator is more loosely constrained from low energy data by
$|\kappa_{Ztc}^R| \leq 0.29$ \cite{tcz}.
While these limits are very interesting, they are indirect limits because
all of the fields in the anomalous operators are virtual, and thus
there could be cancellations between the $Z$-$t$-$c$ (or $\gamma$-$t$-$c$)
operators and other nonstandard physics, and thus the single top
production processes, as direct constraints, are independently valuable.
Studies indicate that from Run~II of the Tevatron,
top decays should provide constraints
of $\Lambda_{gtc} \geq 7.9$ TeV \cite{tcgdecay},
$\kappa_{Ztc}^R \leq 0.3$ \cite{tcz}, 
and will not improve the bounds on
$\Lambda_{\gamma tc}$ from the current $b \ra s \, \gamma$ limit of
about 5 TeV \cite{tcadecay}, unless a new physics signal is found.  
Of course, as we have argued before, it was necessary to
assume a SM $W$-$t$-$b$ interaction in order to use decays to say
anything at all about these operators.  The effect of the $Z$-$t$-$c$
operator to the inclusive $t$-channel production rate is
approximately independent of whether or not the operator is left-handed
or right-handed, and thus in the discussion below we consider
$\kappa_{Ztc}$, which can be taken either as $\kappa_{Ztc}^L$
or $\kappa_{Ztc}^R$.
The effect of these operators on the $t$-channel cross section is
to contribute an additional
0.13 pb at the Tevatron Run~II and 12.6 pb at the LHC, assuming
$\kappa_{Ztc} = 0.29$, and including the 
NLO QCD corrections
for both $t$ and $\bar{t}$ production.
(These should be compared with the SM rates of 2.4 pb and 243 pb, 
at Tevatron and LHC respectively.)

As pointed out above, the constraints
$\kappa_{Ztc}^L, \kappa_{Ztc}^R < 0.05, 0.29$
were obtained by analyzing the low energy data with certain assumptions made
for the underlying theory \cite{tcz}. 
When additional new physics effect is added
to the effect of the $\kappa_{Ztc}$ coupling to the low energy data, it
may turn out that a large cancellation among various sources allow the
size of $\kappa_{Ztc}$ to be at the order of 1. 
Hence, a direct test at
high energy colliders by studying the single top production is necessary
to conclusively determine the coupling $\kappa_{Ztc}$.
For $\kappa_{Ztc} = 1$,
the expected additional single top production rate for
the $W$-gluon mode is 1.56 pb at Run II of the
Tevatron, and 146 pb at the LHC, which in both cases would clearly be
observable as deviations from the SM rates \cite{new}. 
The $\gamma$-$t$-$c$ operator can be studied at a hadron collider
through the reaction $\gamma \, c \ra t$ (where the photon is treated
as a parton inside the proton), though this 
exotic production mechanism suffers from potentially large SM backgrounds.
In contrast, an electron Linear Collider will be better 
suited for this task.

\section{Polarization}
\label{polarization}
\indent \indent

The polarization of top quarks represents another way to probe the
properties of top interactions.  
In the SM, the $W$-$t$-$b$ vertex
is entirely left-handed, which means that the top polarization
information is passed on to the $W$ boson and $b$ quark into which the
top decays.  Since the $W$ interaction with the light fermions
into which it decays is also left-handed, the $W$ polarization information
is thus also reflected in the kinematics of its decay products.
The same weak interaction is also responsible for single top production,
which has the consequence that single tops also show a large degree of
polarization.  The discussion below is based on the SM amplitudes for
top production and decay presented in \cite{dthesis}.

\subsection{The $W^+$ Polarization: The $W$-$t$-$b$ Interaction}
\indent \indent

In order to probe the chiral structure of the $W$-$t$-$b$ interaction,
it is enough to consider the $W$ polarization of top decays.
As was shown in \cite{dthesis}, the left-handed nature of the SM interaction
demands that the produced $W$ bosons be either left-handed or longitudinally
polarized,\footnote{We neglect the tiny mass of the bottom quark,
as is justified given $(m_b / M_W)^2 \sim 3.6 \times 10^{-4}$.}
and predicts the fraction of the longitudinally polarized
$W$ bosons from top decays to be  
\bea
   f_0 = \frac{ m_t^2}{2 \, M_W^2 + m_t^2} \simeq 70\%.
\eea
The degree of $W$ polarization from top decays can be reconstructed by
studying the angle between the charged lepton momentum (defined 
in the $W$ rest frame) and the $W$ momentum defined in the top rest
frame \cite{dthesis}.
Since $t \bar t$ pairs are predominantly produced by QCD interactions
($q \bar q, gg \ra t \bar t$), which conserve parity, 
the top quark is not polarized in its inclusive production,
though there are correlations between the $t$ and $\bar{t}$ spin
at the Tevatron because the dominant production is through a spin 1
gluon \cite{ttbarpol}. 
Hence, it is best to study top decays in $t \bar t$ events to test 
the SM left-handed nature of the $W$-$t$-$b$ coupling by verifying the
fraction of left-handed $W$ bosons from top decays to be $(1-f_0)$.

When probing the $W$-$t$-$b$ interaction from top decays,
the $W$ and $b$ are observed, thus one can be sure that 
it is this interaction that
is responsible for the effect one is seeing, which may not be the case
if there is new physics in single top production.  
Once the chiral structure of the $W$-$t$-$b$ interaction is 
determined, one can
then employ this information to unfold the top decay and reconstruct the
polarization of the top itself, as will be explained below.

\subsection{The Top Polarization}
\indent \indent

Once the chiral structure of the $W$-$t$-$b$ interaction 
has been probed through top decays, and the
SM left-handed structure verified, the top decay products can be
used in order to study the polarization of the produced top quarks
themselves.  As we will see, this can be very useful in determining
what sort of new physics is responsible for an observed deviation in
single top production.  Currently, there are two important bases for
describing the top polarization.  The usual helicity basis measures
the component of top spin along its axis of motion (in the center of
mass frame - because the top mass is large, its helicity is not a Lorentz
invariant quantity).  The so-called ``optimized basis'' 
\cite{optpol} relies on the SM
dynamics responsible for single top production in order to find
a direction (either along the direction of one of the incoming hadrons
or produced jets) which results in a larger degree of polarization for
the top quark.  In the discussion below, we will describe the
modes of single top production in both bases, and analyze the
particular strengths and weaknesses of each.

\begin{figure}[t]
\epsfxsize=6.0in
\centerline{\epsfbox{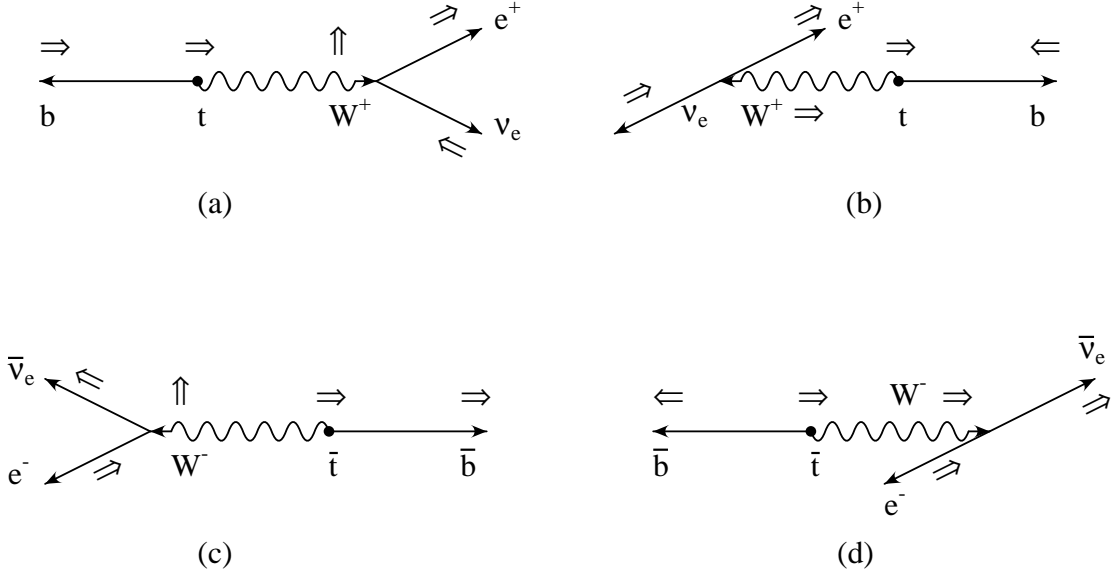}}
\caption{
A diagram indicating schematically the correlation between the
charged lepton ($e^\pm$) from a top decay, and the top spin, in the top
rest frame.  The arrows on the lines indicate the preferred direction
of the momentum in the top rest frame, while the large arrows 
alongside the lines indicate
the preferred direction of polarization.  The figures correspond to
top decay ((a) and (b)) and $\bar{t}$ decay ((c) and (d)) for the
cases when the intermediate $W^\pm$ boson is longitudinally polarized
((a) and (c)) or left-handed ((b) and (d)).  In all cases,
the $e^+$ ($e^-$)
from a $t$ ($\bar{t}$) decay prefers to travel along (against) the
direction of the $t$ ($\bar{t}$) polarization.}
\label{toppolfig}
\end{figure}

Before looking at a particular process or 
basis, it is worth describing how one can determine the top polarization
from its decay products in the decay mode of 
$t \ra W^+ \, b  \ra \ell^+ \, \nu_\ell \, b \,$ \cite{dthesis}.  
A simple heuristic argument based on the
left-handed nature of the $W$ interactions and the conservation of
angular momentum can be made, and is displayed
diagrammatically in Figure~\ref{toppolfig}.  
The analysis is carried out
in the rest frame of the top quark.
When the $W$ boson is longitudinally polarized, it prefers
to move in the same direction as the top spin, 
cf. Figure~\ref{toppolfig}(a). Its decay products
prefer to align along the $W$ polarization, and since the $W$ is boosted
in the direction of the top polarization, the charged lepton again
prefers to move along the top spin axis. 
In the left-handed $W$ case, the fact that the
$b$ quark must be left-handed forces it to move along the direction
of the top polarization, cf. Figure~\ref{toppolfig}(b). 
The $W$ thus moves against this direction.
When the $W$ decays, the charged lepton ($\ell^+$)
must be right-handed, so it prefers
to move against the $W$ direction, in the same direction as the top
polarization.  
Also shown in 
Figures~\ref{toppolfig}(c) and ~\ref{toppolfig}(d),
a similar argument can be made for the 
$\bar{t}$ spin, but in this case the charged lepton prefers to
move against the $\bar{t}$ spin axis.  From this point onward,
we restrict our discussion to top quarks, but it should be clear
how they apply to $\bar{t}$ as well.
The simple angular momentum argument is
reflected in a more detailed computation of the distribution
\cite{topang},
\bea
 \frac{1}{\Gamma} \frac{d \Gamma}{d \cos \theta}
 &=& \frac{1}{2} 
 \left( 1 + \cos \theta \right),
\eea
where $\theta$ is the angle between the
direction of the charged lepton and
the top polarization, in the top rest frame, and
$\Gamma$ is the partial width for a semi-leptonic top decay in the SM.
In principle, one has only to decide on a scheme for relating the top
polarization to some axis, and one can fit the distribution,
\bea
  F(\cos \theta) &=& \frac{A}{2} \left( 1 + \cos \theta \right)
  + \frac{1 - A}{2} \left( 1 - \cos \theta \right) ,
\eea
to determine the degree of polarization ($A$) along this axis.
In practice, there are complications arising from the fact that the
endpoints of the distribution tend to be distorted by the kinematic cuts
required to isolate the signal from the background, and the fact that
in reconstructing the top rest frame, the component of the unobserved
neutrino momentum along the beam axis ($p^{z}_\nu$) is unknown.
One may determine this quantity up to a two-fold ambiguity by
requiring the top decay products to have an invariant mass that
is close to $m_t$.  However, the ambiguity in this procedure will
also have some effect on the distribution, and so careful study is
required.  One can also use the asymmetry between events with
$\cos \theta > 0$ and $\cos \theta < 0$ to characterize the
degree of polarization of the top, which may be helpful if
the data set is limited by poor statistics.

\subsubsection{$W^*$ Production}
\indent \indent

The degree of top polarization in the s-channel
$W^*$ process is straight-forward
to compute in the helicity basis \cite{dthesis}.
Using the CTEQ4M PDF's, we find that at the tree level about 
$75\%$ of the top quarks produced through the
$s$-channel process at the Tevatron are
left-handed, and $76\%$ of them are left-handed at the LHC
\cite{dthesis}.

The optimized basis improves the helicity basis result 
at the Tevatron by noting that
in the SM, the $W^*$ process produces top quarks whose polarization is
always along the direction of the initial anti-quark involved in the
scattering.  At the Tevatron, the vast majority ($\sim 97\%$) of these 
anti-quarks come from the $\bar{p}$ (which has valence anti-quarks).
Thus, one expects that by choosing to measure the top polarization along
the $\bar{p}$ direction in the top rest frame, one can raise the degree
of polarization from $75\%$ to $97\%$.  This represents a large improvement
for Tevatron polarization studies of the $W^*$ process.  However, at the
LHC there are no valence anti-quarks, and thus no optimized basis to
analyze the $W^*$ top polarization. In that case, the helicity basis is
the sensible basis to analyze the polarization of the top quark,
and results in a fair degree of left-handed top production.

\subsubsection{$W$-gluon Fusion}
\indent \indent 

The discussion of polarization in the $W$-gluon fusion process is somewhat
tricky, mostly owing to the fact that the 
detailed kinematics
of this process are sensitive to higher orders of perturbation theory
\cite{tchannel}.
It is clear that the kinematic region described by the process
$q \, b \ra q^\prime \, t$ is the dominant one, but a precise calculation
of the interplay between the $2 \ra 2$ scattering contribution and the
$2 \ra 3$ scattering contribution is still lacking.  Thus, one must be careful
in claiming what
degree of polarization results from a particular basis.

In the helicity basis, the $2 \ra 2$ description has the top
quarks $100\%$ left-handed when
produced from the $u \, b \ra d \, t$ sub-process.  In fact,
at both Tevatron and LHC the $\bar{d} \, b \ra \bar{u} \, t$
sub-process is quite small, and thus the over-all degree of polarization
is about $97\%$.  On the other hand, the $2 \ra 3$ description
shows a degree of polarization that is much lower, and depends 
on the choice of the regularization scheme for the collinear singularity
(e.g., the bottom mass) used in the computation.  This is an
indication that this method of computation is not perturbatively stable.
Thus, it is fair to say that the degree of polarization in the helicity
basis is high, but at the moment no reliable determination is available.
This situation should be improved by including higher order QCD corrections
in $W$-gluon fusion simulation.

The optimized basis once again makes use of the fact that the top
polarization is $100\%$ along the direction of the spectator
anti-quark in the reaction.  At both Tevatron and LHC, this is
dominantly the spectator jet in the final state.  This basis thus
results in a top which is about $96\%$ polarized along the direction
of the spectator jet.  In \cite{optpol}, it was shown that this
basis is not sensitive to the value of the bottom mass, and thus
is perturbatively reliable.  In other words,
higher order QCD effects are unlikely to have a large impact on the
degree of top polarization in the optimized basis.

\subsection{New Physics and Top Polarization}
\indent \indent

As we have seen, new physics may alter the structure of single top
production.  It may be that the new physics effects will reveal
themselves, and tell us something about their nature by causing a
large deviation in one or more of the single top production cross
sections.  In that case one can study the distribution of the top
polarization in order to learn something further about the nature of
the nonstandard production mechanism.

In Section~\ref{extrap}, it was demonstrated that either a
charged scalar top-pion or $W^\prime$ gauge boson can have
a substantial effect on single top production in the $s$-channel
mode.  Assuming for the moment that such a deviation has been
observed, one can then use the top polarization in order to narrow
down the class of models responsible for such an effect.  The $W^\prime$
boson couples to the left-handed top and bottom quarks, and thus an
analysis of the resulting top polarization will be similar to that of the SM
prediction.  Namely, the helicity basis will show about $75\%$ of the tops
to be left-handed ($75\%$ at the LHC), though the specific numbers
show a mild sensitivity to the $W^\prime$ mass, and the optimized basis will
show about $97\%$ at the Tevatron.  However, the $\pi^\pm$ has a right-handed
interaction, completely at odds to the SM.  In fact, there is another
difference between the $W^\prime$ and the $\pi^\pm$ that is also very
important.  Like the SM $W$ boson, the $W^\prime$ is a vector particle,
and thus carries angular momentum information between the initial state
and final state in the $s$-channel process.  However, the $\pi^\pm$,
as a scalar particle, does not carry such information.  
Thus, the optimized
basis, which relies on the correlation between 
top spin and the initial $\bar{d}$
momentum fails to apply to a scalar production mechanism, and if one
were to use it to analyze the polarization of the top coming from this
type of new physics effect, one would come to the wrong conclusion that
the produced tops were unpolarized.  On the other hand, in the
helicity basis the top quarks produced from the $\pi^\pm$ show
very close to $100\%$ right-handed polarization.  
This demonstrates the
utility of using {\em both} bases.  If there is new physics in single top
production, not only is it unclear at the outset which basis 
will show a larger
degree of polarization, but we can use them together to distinguish
a vector from a scalar exchange, thus learning about the nature of
the new particle without directly observing it.

Study of polarization can also be useful in disentangling the operators
in the effective Lagrangian in Equations~(\ref{ewcleq1}) 
and (\ref{ewcleq2}).  As we saw, those operators have left-handed and
right-handed versions, and thus the distribution of top polarizations
will depend on the relative strength of the two.  Thus, by studying top
polarization, one could begin to disentangle the chiral structure of
the operator responsible for a deviation in single top production,
giving further insight into the nature of the full theory that 
accurately describes higher energies.

\section{Conclusions}
\label{conclusions}
\indent \indent

Having gone over in detail the physics one can probe with single top
production, it is worth summarizing what we have learned and examining
how one can use the different top quark observables to extract information
about the top that maximizes the available information.  In the preceding
sections we have seen that single top production allows one to measure the
magnitude of the top's weak interactions 
(in contrast to top decays).  The three
modes of single top production 
(the s-channel $W^*$, the t-channel $W$-gluon fusion, and the \tw modes)
are sensitive to different types of
new physics.  All three modes are sensitive to modification of the
$W$-$t$-$b$
interaction, with the $t \, W^-$ mode distinguished by the fact
that it is rather insensitive to most types of new physics.
The $s$-channel mode is sensitive to certain types of additional (heavy)
particles.  And the $t$-channel mode is sensitive to physics which modifies
the top decay properties, in particular to FCNC interactions.
In this light, it is rather unfortunate that the $t \, W^-$ mode is
so small at the Tevatron that it is not likely to be useful there,
as it can allow one to measure the strength of the $W$-$t$-$b$ vertex,
which would be a good first step in disentangling the information from
the $s$- and $t$-channel modes.

\begin{figure}[p]
\epsfxsize=6.0in
\centerline{\epsfbox{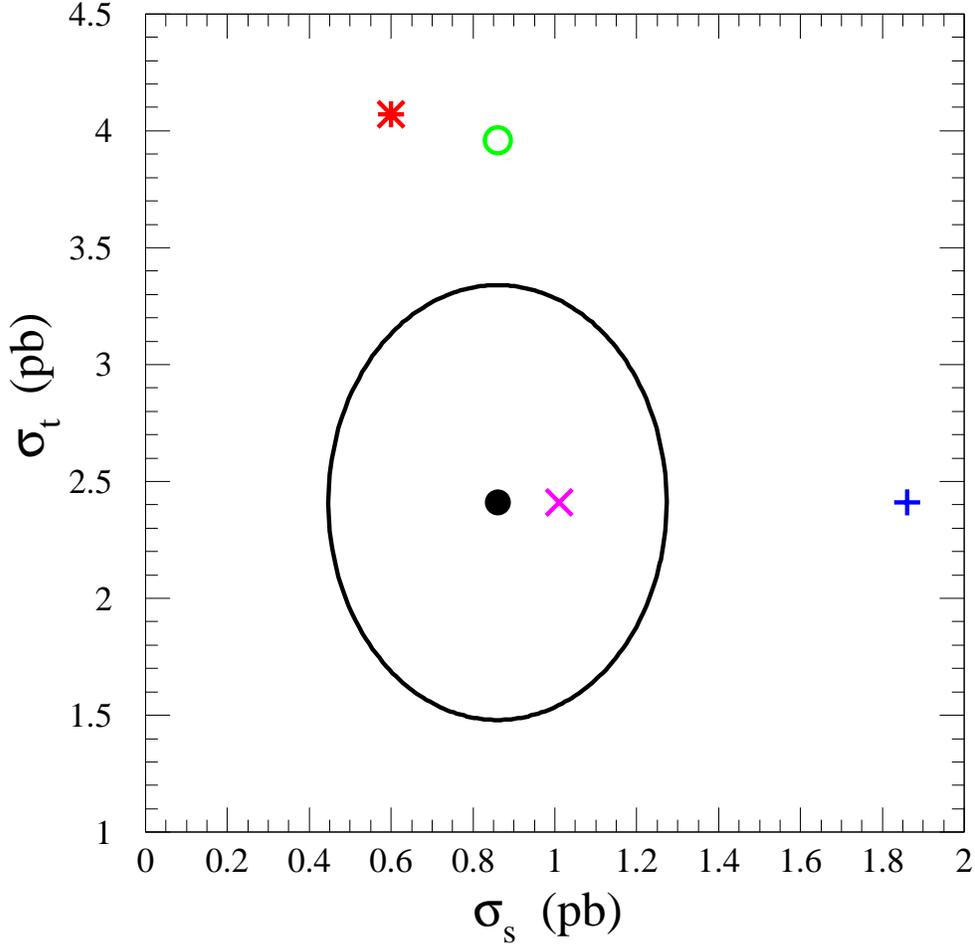}}
\caption{The location of the Tevatron SM point
(the solid circle) in the $\sigma_s$-$\sigma_t$ plane,
and the $3 \sigma$ theoretical deviation
curve.  Also shown are the points for
the top-flavor model (with $M_Z^{\prime} = 1$ TeV
and $\sin^2 \phi = 0.05$) as the X, the
FCNC $Z$-$t$-$c$ vertex
($|\kappa_{Ztc}|=1$) as the open circle,
a model with a charged top-pion ($m_{\pi^{\pm}} =$
250 GeV and $t_R$-$c_R$ mixing of $\sim 20\%$) as the cross,
and a four quark generation scenario
with $|V_{ts}| = 0.55$ and $|V_{tb}| = 0.835$ as the asterisk.
All cross sections sum the $t$ and $\bar{t}$ rates.}
\label{sigmaplanefig}
\end{figure}

\begin{figure}[p]
\epsfxsize=6.0in
\centerline{\epsfbox{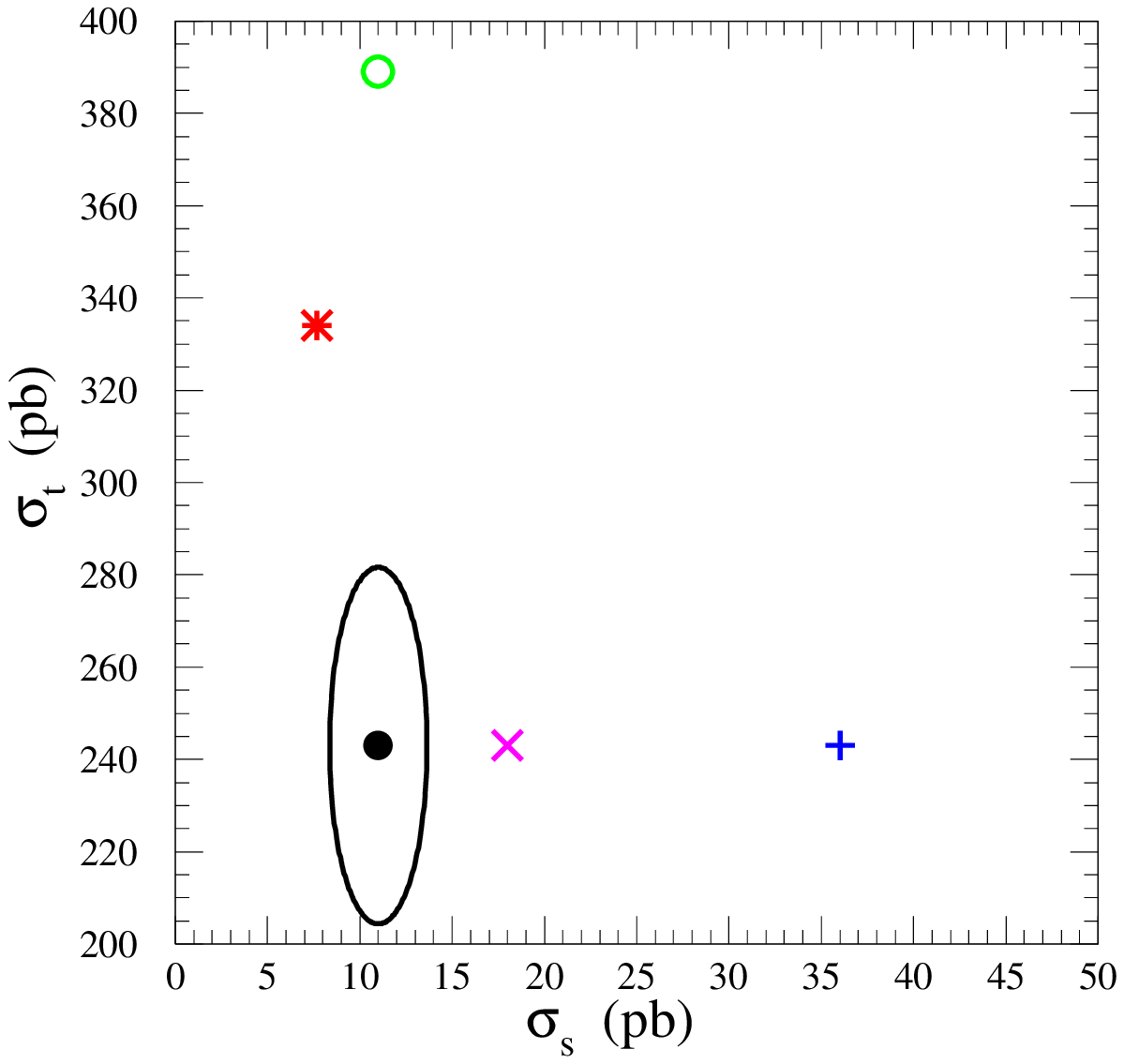}}
\caption{The location of the LHC SM point
(the solid circle) in the $\sigma_s$-$\sigma_t$ plane,
and the $3 \sigma$ theoretical deviation
curve.  Also shown are the points for
the top-flavor model (with $M_Z^{\prime} = 1$ TeV
and $\sin^2 \phi = 0.05$) as the X, the
FCNC $Z$-$t$-$c$ vertex
($|\kappa_{Ztc}|=1$) as the open circle,
a model with a charged top-pion ($m_{\pi^{\pm}} =$
450 GeV and $t_R$-$c_R$ mixing of $\sim 20\%$) as the cross,
and a four quark generation scenario
with $|V_{ts}| = 0.55$ and $|V_{tb}| = 0.835$ as the asterisk.
All cross sections sum the $t$ and $\bar{t}$ rates.}
\label{lhcplanefig}
\end{figure}

Without the $t \, W^-$ mode, one will most likely have to study the
correlation of the $s$- and $t$- channel rates in the plane of 
$\sigma_s-\sigma_t$ in order to attempt to understand if a new
physics effect is present, and how one should interpret it if it is
observed.  
($\sigma_s$ and $\sigma_t$ are the cross sections of $W^*$ and 
$W$-gluon processes, respectively.)
In Figure~\ref{sigmaplanefig} we show this plane for Run II of the
Tevatron,
including the SM point (with the contour of $3 \, \sigma$
theoretical uncertainty deviation around it)
and the points from the the top-flavor model
(with $M_{Z^\prime} = 900$ GeV and $\sin^2 \phi = 0.05$), 
the top-color model with a charged top-pion (with mass
$m_\pi^\pm = 250$ GeV and $t_R$-$c_R$ mixing of $20\%$),
a FCNC $Z$-$t$-$c$ operator (with $|\kappa_{Ztc}|=1$ 
and $\phi_{Ztc}^R = \phi_{Ztc}^L = 0$),
and a large $V_{ts}$ under the four quark generation scenario
with $|V_{ts}| = 0.55$ and $|V_{tb}| = 0.835$.
A similar plot is given in Figure~\ref{lhcplanefig} for the LHC.
This illustrates how to use the knowledge we have about the sensitivity
of the $W^*$ and $W$-gluon fusion modes to find a likely explanation for
a new physics effect.  A deviation in $\sigma_s$ that is not
also reflected in $\sigma_t$ is most likely due to the effect of
nonstandard particles.  A deviation in $\sigma_t$ that
is not also seen in $\sigma_s$ is likely from a FCNC.  A deviation
that is comparable in both rates is most likely from a modification
of the $W$-$t$-$b$ interaction.  In the very least, if the SM is a
sufficient description of single top production, the fact that the
two rates are consistent will allow one to use them to extract
$V_{tb}$ with confidence that new physics is not distorting the
measurement.

Additional information is provided by polarization information.
By studying the $W$ polarization from top decays, one learns
about the nature of the $W$-$t$-$b$ interaction.  By studying the
top polarization, in both the helicity and optimized bases,
one can learn more about the chiral structure of nonstandard top
interactions, either by probing the chiral structure of the
interactions, or even the scalar/vector nature of a virtual
particle participating in single top production.

The large top mass seems to be a hint that the
mechanism of the EWSB may be more evident in studies of top than in
other observables, and thus the Tevatron Run~II and the LHC are
exciting opportunities to probe the nature of the symmetry breaking.
We have seen that by using the three modes of single top production
together, along with studies of polarization in top decays and
in single top production, one can assemble a coherent picture of the
properties of the top.  These observables a sensitive to different
kinds of new physics, and thus when considered together can provide
a probe of the nature of nonstandard physics manifest in the top 
sector, or can demonstrate the validity of the SM picture of how the
top should behave.   

\section{Acknowledgements}
\indent \indent

T. Tait has benefitted from discussions with E.L. Berger, B.W. Harris, 
T. Lecompte, E. Malkawi, Z. Sullivan, and S. Willenbrock.
We thank H.--J. He and W. Repko for collaboration on the
associated 
production of a scalar boson with a single top quark.
Work at Argonne National Lab 
is supported in part by the DOE under contract W-31-109-ENG-38.  
CPY is supported in part by the NSF under the grant PHY-9802564.  



\newpage
\begin{thebibliography}{99}

\bibitem{topdisc}
F. Abe {\it et al}, CDF Collaboration, Phys. Rev. Lett. {\bf 74}, 2626
(1995);\\
S. Abachi, {\it et al}, D0 Collaboration, Phys. Rev. Lett. {\bf 74},
2632 (1995).

\bibitem{topcondensate}
W.A.~Bardeen, C.T.~Hill, and M.~Linder, Phys. Rev. {\bf D41}, 1647 (1990).

\bibitem{topcolor}
C.T.~Hill, Phys. Lett. {\bf B345}, 483 (1995); \\
B.A. Dobrescu, C.T. Hill, Phys. Rev. Lett. {\bf 81}, 2634 (1998); \\
R.S. Chivukula, B.A. Dobrescu, H. Georgi, C. T. Hill,
Phys. Rev. {\bf D59}, 075003 (1999); \\
M.B. Popovic, E.H. Simmons, Phys.\ Rev.\  {\bf D62}, 035002 (2000); \\ 
H.--C. Cheng, B.A. Dobrescu, and C.T. Hill, hep-ph/9912343; \\
H. Collins, A. Grant, and H. Georgi 
Phys. Rev. {\bf D61}, 055002 (2000); \\
H. Georgi, and A.K. Grant, hep-ph/0006050; \\
A.~Aranda and C.D.~Carone, hep-ph/0007020.

\bibitem{topflavor}
R.S. Chivukula, E.H. Simmons, J. Terning, 
Phys. Lett. {\bf B346}, 284 (1995); \\ 
E. Malkawi, Tim Tait, and C.--P. Yuan, Phys. Lett. {\bf B385}, 304 (1996); \\
D.J. Muller, S. Nandi, Phys. Lett. {\bf B383}, 345 (1996); \\
J.C. Lee, K.Y. Lee, and J.K. Kim, Phys. Lett. {\bf B424}, 133 (1998); \\
K.R.~Lynch, E.H.~Simmons, M.~Narain and S.~Mrenna, hep-ph/0007286.

\bibitem{tf1}
E. Malkawi and C.--P. Yuan, Phys. Rev. {\bf D61}, 015007 (2000).

\bibitem{tf2}
H.--J. He, Tim Tait, C.--P. Yuan, Phys. Rev. {\bf D62}, 011702 (2000).

\bibitem{mythesis}
For a comprehensive introduction,\\
Tim~M.P.~Tait, Ph.D. thesis, Michigan State University, 
hep-ph/9907462 (1999).

\bibitem{LHCreport}
M.~Beneke {\it et al.}, ``Top quark physics'', hep-ph/0003033.

\bibitem{adlernelson}
C.A.~Nelson and L.J.~Adler Jr., hep-ph/0006342; hep-ph/0007086.

\bibitem{mrenna}
S.~Mrenna and C.--P.~Yuan, Phys. Rev. {\bf D46}, 1007 (1992).

\bibitem{nlcztt}
J. Bagger, {\it et al.},
``The case for a 500-GeV $e^+ e^-$ linear collider'', hep-ex/0007022.

\bibitem{tchannel}
S. Dawson, Nucl. Phys. {\bf B249}, 42 (1985);\\
S. Willenbrock and D. Dicus, Phys. Rev. {\bf D34}, 155 (1986); \\
C.--P. Yuan, Phys. Rev. {\bf D41}, 42 (1990);
CCAST Symposium 1993, 259 (1993);
Valencia Elem. Part. Phys. 1995, 148 (1995);
{\it 5th Mexican Workshop of Particles and Fields}, Puebla, Mexico
(1995); \\
R.K. Ellis, and S. Parke Phys. Rev. {\bf D46}, 3785 (1992);\\
G. Bordes, and B. van Eijk, Z. Phys. {\bf C 57}, 81 (1993);
Nucl. Phys. {\bf B435}, 23 (1995); \\
T. Stelzer, Z. Sullivan, and S. Willenbrock, 
Phys. Rev. {\bf D56}, 5919 (1997).

\bibitem{dthesis}
D.O. Carlson and C.--P. Yuan, Phys. Lett. {\bf B306}, 386 (1993); \\
D.O. Carlson, Ph.D. Thesis, Michigan State University (1995).

\bibitem{newt}
T. Stelzer, Z. Sullivan, and S. Willenbrock,
Phys. Rev. {\bf D58}, 094021 (1998).

\bibitem{cteq4}
CTEQ Collaboration: H. Lai, J. Huston, S. Kuhlmann,
F. Olness, J. Owens, D. Sopher, W.--K. Tung, and H. Weerts,
Phys. Rev. {\bf D 55}, 1280 (1997).

\bibitem{schannel}
S. Cortese and R. Petronzio, Phys. Lett. {\bf 253B}, 494 (1991);\\
T. Stelzer and S. Willenbrock, Phys. Lett. {\bf B357}, 125 (1995);\\
M.C. Smith and S. Willenbrock, Phys. Rev. {\bf D54}, 6696 (1996); \\
S. Mrenna and C.--P. Yuan, Phys. Lett. {\bf B416}, 200 (1998).

\bibitem{new}
D.O. Carlson, and C.--P. Yuan, Particle Phys. \& Phen. 1995, 172 (1995); \\
Tim Tait and C.--P. Yuan, hep-ph/9710372.

\bibitem{tw}
G. Ladinsky, and C.--P. Yuan, Phys. Rev. {\bf D43}, 789 (1991);\\
S. Moretti, Phys. Rev. {\bf D56}, 7427 (1997); \\
A.P. Heinson, A.S. Belyaev, and E.E. Boos, 
Phys. Rev. {\bf D56}, 3114 (1997); \\
A.S. Belyaev, E.E. Boos, and L.V. Dudko, 
Phys. Rev. {\bf D59}, 075001 (1999); \\
A.S. Belyaev, E. Boos, hep-ph/0003260.

\bibitem{mytw}
Tim M. P. Tait, Phys. Rev. {\bf D61}, 034001 (2000).

\bibitem{cpviol}
D. Atwood, S. Bar-Shalom, G. Eilam, and A. Soni, hep-ph/0006032.

\bibitem{susy}
For a review,
S.~Abel {\it et al.},
``Report of the SUGRA working group for run II of the Tevatron'',
hep-ph/0003154.

\bibitem{genstnp}
C.--S. Li, R. Oakes, and J.--M. Yang, Phys. Rev. {\bf D55}, 1672 (1997); 
Phys. Rev. {\bf D55}, 5780 (1997); \\
C.--S. Li, R. Oakes, J.--M. Yang, H.--Y. Zhou,
Phys. Rev. {\bf D57}, 2009 (1998); \\
S. Bar-Shalom, D. Atwood, A. Soni, Phys. Rev. {\bf D57}, 1495 (1998).

\bibitem{pdg}
C. Caso et al, The European Physical Journal {\bf C3}, 1 (1998) 
and 1999 off-year partial update for the 2000 edition available on 
the PDG WWW pages (URL: http://pdg.lbl.gov/). 

\bibitem{tevgut}
K. R. Dienes, E. Dudas, and T. Gherghetta, Phys. Lett. {\bf B436}, 55 (1998);
Nucl. Phys. {\bf B537}, 47 (1999).

\bibitem{leptoquark}
K.~Agashe and M.~Graesser, Phys.\ Rev.\  {\bf D54}, 4445 (1996).

\bibitem{gmsb}
For a review,
C. Kolda, Nucl. Phys. Proc. Suppl. {\bf 62}, 266 (1998).

\bibitem{CDFvtb}
J. Incandela (CDF Collaboration), Fermilab-Conf-95/237-E (1995).

\bibitem{flavoron}
G.~Burdman, R.S.~Chivukula and N.~Evans, hep-ph/0005098.

\bibitem{es}
E.H. Simmons, Phys. Rev. {\bf D55}, 5494 (1997).

\bibitem{singletxd}
A.~Datta, P.J.~O'Donnell, Z.H.~Lin, X.~Zhang and T.~Huang,
Phys.\ Lett.\  {\bf B483}, 203 (2000).

\bibitem{toppion}
H.--J. He and C.--P. Yuan, 
Phys.\ Rev.\ Lett.\  {\bf 83}, 28 (1999).

\bibitem{nlotoppion}
C. Balazs, H.--J. He and C.--P. Yuan, Phys.\ Rev.\  {\bf D60}, 114001 (1999).

\bibitem{toppi0}
G.~Burdman, Phys.\ Rev.\ Lett.\  {\bf 83}, 2888 (1999).

\bibitem{tcst}
P. Baringer, P. Jain, D.W. McKay, and L. Smith,
Phys. Rev. {\bf D56}, 2914 (1997).

\bibitem{rpviol}
R. Oakes, K. Whisnant, J.--M. Yang, B.--L. Young, X. Zhang,
Phys. Rev. {\bf D57}, 534 (1998).

\bibitem{edbrizack}
E.L. Berger, B.W. Harris, and Z. Sullivan,
Phys. Rev. Lett. {\bf 83}, 4472 (1999).

\bibitem{hsinglet}
J.L. Diaz-Cruz, M.A. Perez, and J.J. Toscano, 
Phys. Lett. {\bf B398}, 347 (1997).

\bibitem{ewcl}
R. Peccei and X. Zhang, Nucl. Phys. {\bf B337}, 269 (1990);\\
E. Malkawi and C.-P. Yuan, Phys. Rev. {\bf D50}, 4462 (1994).

\bibitem{nda}
A. Manohar and H. Georgi, Nucl. Phys. {\bf B234}, 189 (1984).

\bibitem{nonlinear}
R.~D.~Peccei and X.~Zhang, Nucl.\ Phys.\  {\bf B337}, 269 (1990);
R.~D.~Peccei, S.~Peris and X.~Zhang, Nucl.\ Phys.\  {\bf B349}, 305 (1991).

\bibitem{ehabewcl}
D. Carlson, E. Malkawi, and C.--P. Yuan, Phys. Lett. {\bf B337}, 145 (1994); \\
E. Malkawi, and C.--P. Yuan, Phys. Rev. {\bf D50}, 4462 (1994); \\
E. Malkawi, and C.--P. Yuan, Phys. Rev. {\bf D52}, 472 (1995). 

\bibitem{dim5}
F. Larios and C.--P. Yuan, Phys. Rev. {\bf D55}, 7218 (1997); \\
F. Larios, Tim Tait, and C.--P. Yuan, Phys. Rev. {\bf D57}, 3106 (1998); \\
T.~Han and J.L.~Hewett, Phys.\ Rev.\  {\bf D60}, 074015 (1999); \\
E.R. Morales and M.E. Peskin, hep-ph/9909383; \\
T. Han, Y.J. Kim, A. Likhoded, and G. Valencia, hep-ph/0005306.

\bibitem{ttbarpol}
G.~Mahlon and S.~Parke, Phys.\ Rev.\  {\bf D53}, 4886 (1996);
Phys.\ Lett.\  {\bf B411}, 173 (1997).

\bibitem{cleo}
T. Skwarnicki, talk at ICHEP98, Vancouver, Canada (1998); \\
M. Alam et al, CLEO collaboration, Phys. Rev. Lett. {\bf 74}, 2885 (1998).

\bibitem{kappaR}
K. Chetyrkin, M. Misiak, and M. Munz, Phys. Lett. {\bf B400}, 206 (1997); \\
F. Larios, M.A. P\'erez, and C.--P. Yuan, 
Phys.\ Lett.\  {\bf B457}, 334 (1999).

\bibitem{valencia} A. Abd El-Hady and G. Valencia,
Phys. Lett. {\bf B414}, 173 (1997).

\bibitem{ztprod}
F. del Aguila, J.A. Aguilar-Saavedra, and L. Ametller, 
Phys Lett. {\bf B462}, 310 (1999); \\
F. del Aguila, and J.A. Aguilar-Saavedra,
Nucl. Phys. {\bf B576}, 56 (2000).

\bibitem{tcgmenehab}
E. Malkawi and Tim Tait, Phys. Rev. {\bf D54}, 5758 (1996).

\bibitem{tcgother}
M. Hosch, K. Whisnant, and B.--L. Young,
Phys. Rev. {\bf D56}, 5725 (1997); \\
T. Han, M. Hosch, K. Whisnant, B.--L. Young, and X. Zhang,
Phys. Rev. {\bf D58}, 073008 (1998).

\bibitem{tcgdecay}
T. Han, K. Whisnant, B.--L. Young, X. Zhang,
Phys. Lett. {\bf B385}, 311 (1996); \\
Tim Tait and C.--P. Yuan, Phys. Rev. {\bf D55}, 7300 (1997).

\bibitem{tcadecay}
T.~Han, K.~Whisnant, B.--L.~Young and X.~Zhang,
Phys.\ Rev.\  {\bf D55}, 7241 (1997).

\bibitem{tcz}
T. Han, R.D. Peccei, and X. Zhang, 
Nucl. Phys. {\bf B454}, 527 (1995).

\bibitem{topang}
M. Klein, H. Pietschmann, and H. Rupertsberger, 
Phys. Lett. {\bf B153}, 341 (1985); \\
I. Bigi, Y. Dokshitzer, V. A. Khoze, J. K\"uhn, and P. Zerwas,
Phys. Lett. {\bf 181B}, 157 (1986); \\
M. Jezabek and J. K\"uhn, Phys. Lett. {\bf B329}, 317 (1994).

\bibitem{optpol}
G. Mahlon and S. Parke, Phys. Rev. {\bf D55}, 7249 (1997); \\
G. Mahlon and S. Parke, Phys. Lett. {\bf B476}, 323 (2000).

\end{thebibliography}
\end{document}